\documentclass[11pt,letterpaper,draftcls,oneside,peerreview,onecolumn]{IEEEtran}
\usepackage{amsmath}
\usepackage{amsfonts}
\usepackage{amsmath}
\usepackage{graphicx}
\usepackage{amssymb}
\usepackage{booktabs}
\usepackage{cite}
\usepackage{url}
\usepackage{balance}
\usepackage[OT1,T1]{fontenc}
\usepackage{setspace}
\usepackage{color}
\newenvironment{proof}[1][Proof]{\noindent\textbf{#1.} }{\ \rule{0.5em}{0.5em}}

\newcommand{\undersmile}{\mathrel{\lower6pt\hbox{$\smile$}}}

\newcommand{\qh}{{\bf h}}

\newcommand{\qn}{{\bf n}}

\newcommand{\qr}{{\bf r}}

\newcommand{\qt}{{\bf t}}
\newcommand{\qu}{{\bf u}}
\newcommand{\qv}{{\bf v}}
\newcommand{\qw}{{\bf w}}
\newcommand{\qx}{{\bf x}}

\newcommand{\qB}{{\bf B}}

\newcommand{\qD}{{\bf D}}
\newcommand{\qE}{{\bf E}}

\newcommand{\qH}{{\bf H}}
\newcommand{\qI}{{\bf I}}

\newcommand{\qW}{{\bf W}}
\newcommand{\qX}{{\bf X}}

\newcommand{\qzero}{{\bf 0}}

\newcommand{\tr}{\mbox{Tr}}

\newcommand{\be}{\begin{equation}} \newcommand{\ee}{\end{equation}}
\newcommand{\bea}{\begin{eqnarray}} \newcommand{\eea}{\end{eqnarray}}

\begin{document}

\newcounter{MYtempeqncnt}

\title{Low-complexity End-to-End Performance Optimization in MIMO Full-Duplex Relay Systems}
\author{Himal A. Suraweera, Ioannis Krikidis, Gan Zheng, Chau Yuen, and Peter J. Smith
\thanks{\scriptsize H. A. Suraweera is with the Department of Electrical and Electronic Engineering, University of Peradeniya, Peradeniya 20400, Sri Lanka (e-mail: himal@ee.pdn.ac.lk)}
\thanks{\scriptsize H. A. Suraweera and C. Yuen are with the Singapore University of Technology and Design, 20 Dover Drive, Singapore 138682 (e-mail: \{himalsuraweera, yuenchau\}@sutd.edu.sg)}
\thanks{\scriptsize I. Krikidis is with the Department of Electrical \& Computer Engineering, University of Cyprus, Nicosia 1678, Cyprus (e-mail: krikidis@ucy.ac.cy)}
\thanks{\scriptsize G. Zheng is with the Interdisciplinary Centre for Security, Reliability and Trust (SnT), University of Luxembourg, 4 rue Alphonse Weicker, L-2721 Luxembourg (e-mail: gan.zheng@uni.lu)}
\thanks{\scriptsize P. J. Smith is with the Department of Electrical and Computer Engineering, The University of Canterbury, Private Bag 4800, Christchurch, New Zealand (email: peter.smith@canterbury.ac.nz)}
\thanks{\scriptsize This work was presented in part at the IEEE Intl. Conf. Commun. (ICC 2013), Budapest, Hungry, June 2013.}}
\maketitle

\vspace{-0.5cm}

\begin{abstract}
In this paper,  we deal with the deployment of full-duplex relaying in amplify-and-forward (AF) cooperative networks with  multiple-antenna terminals. In contrast to previous studies, which focus on the spatial mitigation of the loopback interference (LI) at the relay node, a joint precoding/decoding design that maximizes the end-to-end (e2e) performance is investigated. The proposed precoding incorporates rank-1 zero-forcing (ZF) LI suppression at the relay node and is derived in closed-form by solving appropriate optimization problems. In order to further reduce system complexity, the antenna selection (AS) problem for full-duplex AF cooperative systems is discussed. We investigate different AS schemes to select a single transmit antenna at both the source and the relay, as well as a single receive antenna at both the relay and the destination. To facilitate comparison, exact outage probability expressions and asymptotic approximations of the proposed AS schemes are provided. In order to overcome zero-diversity effects associated with the AS operation, a simple power allocation scheme at the relay node is also investigated and its optimal value is analytically derived. Numerical and simulation results show that the joint ZF-based precoding significantly improves e2e performance, while AS schemes are efficient solutions for scenarios with strict computational constraints.
\end{abstract}


\section{Introduction}
Cooperative communications with relaying is a promising solution to extend the network coverage and ensure higher throughputs and quality-of-service (QoS). Relaying techniques can be classified as either half-duplex or full-duplex \cite{KRI}. In order to complete the relaying operation, half-duplex relaying requires two orthogonal channels and the associated bandwidth loss recovery has been an active research area for several years. With full-duplex relaying, the relay node receives and transmits simultaneously on the same channel and therefore utilizes the spectrum resources more efficiently \cite{BLISS,BLISS1}. However, the main limitation in full-duplex operation is the loopback interference (LI) (also known in the literature as the loopback self-interference) due to signal leakage from the relay's output to the input at the reception side~\cite{RII3,RII2,RII1,DAY}. Specifically, the main drawback of full-duplex operation is the large power differential between the LI generated by the full-duplex terminal and the received signal of interest coming from a distant source. The large LI spans most of the dynamic range of the analog-to-digital converter at the receiver side and thus its mitigation is critical for the implementation of full-duplex operation. In modern communication systems such as WiFi, Bluetooth, and Femtocells, the transmission power and the distance between communicating devices has been decreased. This important architectural modification decreases the power differential between the two received signals. This attribute, combined with the high computation capabilities of modern terminals, significantly facilitates the implementation of  the full-duplex radio technology \cite{DUARTE1,DUARTE2,JAIN}.

In the literature, the combination of multiple-input multiple-output (MIMO) techniques with relaying has been invoked to further enhance the communication performance \cite{FAN,MO}. While most work has focused on MIMO half-duplex relaying, recent work has also considered MIMO full-duplex relaying. MIMO provides an effective means to suppress the LI in the spatial domain \cite{RII1,SUNG,LIOLIOU}. With multiple transmit or receive antennas at the full-duplex relay, precoding at the transmitter and decoding at the receiver can be jointly optimized to mitigate the LI effects. Zero forcing (ZF) and minimum mean square error (MMSE) are two widely adopted criteria in the literature for the precoding and decoding design \cite{COMM_MAG_ZHANG}. ZF aims to completely null out undesired interference and provides an interference-free channel. Although ZF normally results in sub-optimal solutions, its performance is nearly optimal in the high signal-to-noise ratio (SNR) regime. MMSE is an improved precoder/decoder design compared to ZF, which takes into account the background noise. The MMSE-based precoder has a more complicated structure but it can improve the achievable QoS. Due to the implementation simplicity and the efficiency in the high SNR regime, ZF becomes a useful design criterion to completely cancel the LI and break the closed-loop between the relay input and output.

Assuming there is no closed-loop processing delay, the optimal precoding matrix for a full-duplex amplify-and-forward (AF) relay that maximizes the mutual information under an average power constraint is studied in \cite{Kang-09}. In this case, the design approach and the resulting precoding solution are similar to the half-duplex case. The joint precoding and decoding design for a full-duplex relay is studied in \cite{Riihonen-Spatial-09,RII1}, where both ZF solutions and MMSE solutions are discussed. Notice that the ZF solution used in \cite{Riihonen-Spatial-09,RII1} and most early works uses a conventional approach based on the singular value decomposition of the loopback self-interference channel. The main drawback of this approach is that the ZF solution only exists given that the numbers of antennas at the source, full-duplex relay and the receiver satisfy a certain condition. In order to overcome this limitation, \cite{LIOLIOU} adopts an alternative criterion and proposes to maximize the signal-to-interference ratios between the power of the useful signal to the power of LI at the relay input and output, respectively. Conventional ZF precoding and decoding are chosen via the singular vectors of the LI channels, however, this design does not take into account the other channels and the end-to-end (e2e) performance. In \cite{CHOI_EL}, a joint design of ZF precoding and decoding is proposed to fully null out the LI at the relay, taking into account the source-relay and relay-destination channels. A simple approach is studied in \cite{Park-null-12}, where an iterative algorithm that jointly optimizes the precoding and decoding vectors in respect of the e2e performance, is investigated.

Most of the work in the literature does not deal with the joint optimization of the precoding and decoding process, even for scenarios with multiple antennas at the terminals. Hence, the focus has been restricted to full duplex relay processing which has led to strictly suboptimal e2e performance. Furthermore, the available ZF-based solutions which do aim to optimize e2e performance are not given in closed-form. Hence, in this paper, we consider a general case where each terminal can have arbitrary multiple antennas and we jointly design precoding and decoding at the source, the relay and the destination in order to maximize the
achievable rate. For simplicity, a single data stream is transmitted and ZF criteria are used by the full-duplex relay to handle the LI. We give the closed-form precoder/decoder solutions for transmit and receive ZF schemes. Furthermore, the diversity orders are derived for the different schemes.

In addition, we also propose several low-complexity antenna selection (AS) schemes\footnote{We follow the footsteps of recent work such as \cite{RII1} and investigate the performance of AS since both precoding/decoding and AS schemes belong to the general category of MIMO spatial suppression techniques. ZF precoding/decoding designs and different AS schemes studied in this paper eliminate/mitigate the effect of LI respectively, and hence offer different performance/complexity tradeoff choices to a system designer. Moreover, AS can be viewed as a special case of precoding where the beamforming vector only contains a single non zero unit element whose entry depends on the selected transmit/receive antenna.} for MIMO full-duplex relaying and analyze the outage probability of each scheme. The complexity of implementing MIMO systems can be significantly decreased with AS, which employs fewer radio frequency chains than antenna elements and then connects the chains to the best available antenna element~\cite{MOLISCH}. Some limited work on AS in full-duplex relay systems can be found in \cite{RII1,SUNG}. In \cite{RII1}, several spatial LI suppression techniques based on antenna sub-set selection and joint transmit/receive beam selection have been investigated. In \cite{SUNG}, several low complexity antenna sub-set selection schemes have been proposed with the objective to suppress LI at the relay's transmit side. However, a basic limitation of the current work is that AS is used only to achieve LI suppression. On the other hand, from a system performance standpoint, it is important to deploy MIMO AS techniques such that the e2e signal-to-interference noise ratio (SINR) at the destination is maximized.

The performance of AS in half-duplex relay systems is a mature topic and well studied, see for e.g., \cite{YANG,GAYAN,HIMAL,HWCNC10}. On the other hand, to the best of authors' knowledge, the current paper is the first to analytically investigate the AS performance for full-duplex relay systems. Moreover, our analysis presents new results in addition to earlier work such as \cite{RII200,RII300} where the outage probability of single antenna full-duplex systems have been studied. Specifically, we select single transmit antennas at the source and the relay, respectively, and single receive antennas at the relay and the destination, respectively. The performance of the aforementioned system set-up with different AS schemes is quantified by deriving exact, and asymptotic outage probability expressions. The asymptotic expressions illuminate the network performance by revealing the comparative performances of the AS schemes in terms of the system and channel parameters. Furthermore, in order to eliminate the zero-diversity behavior of the full-duplex relaying due to the LI, we propose a new simple power allocation scheme\footnote{It should be noted that due to the influence of LI, power adaptation (or ``gain control'' \cite{RII300}) is an important issue for full-duplex AF relaying.} at the relay, which only involves a single parameter optimization. We also present optimal values for this parameter to minimize the outage probability from a diversity perspective. These closed-form expressions are in the form of fractions of the number of source/relay/destination antennas and reveal the spatial degrees of freedom offered by each AS scheme. Moreover, these values can be calculated directly once a particular system configuration is decided.

The main contributions of this paper are twofold.
\begin{itemize}
\item
A low complexity joint precoding/decoding design for e2e SNR maximization is proposed. Specifically, based on ZF loopback self-interference suppression, receive/transmit beamforming vectors at the relay are designed. Closed-form solutions for the scheme's outage probability as well as high SNR simple expressions are derived. Our analysis clearly reveals insights on system performance and shows the impact on the achieved diversity order.
\item Several AS schemes are proposed including the optimal AS scheme that maximizes the e2e SNR at the destination and various sub-optimal AS schemes. In order to eliminate the zero diversity behavior in such full-duplex MIMO systems, we propose a simple power allocation method at the relay. The outage performance of the AS schemes are analytically investigated. Using the derived high SINR outage approximations, we also investigate the optimal power allocation coefficient values.
\end{itemize}

The rest of the paper is organized as follows. Section II presents the overall MIMO system model. Sections III and IV present the joint precoding/decoding designs and AS schemes, respectively. The outage probability of the precoding and AS schemes is analyzed in Section V and numerical results are given in Section VI. Finally, Section VII concludes the paper and summarizes several key findings.

\noindent \emph{Notation}: The lowercase and uppercase boldface letters (e.g., $\qx$ and $\qX$) indicate column vectors and matrices, respectively. $\qI$ is the identity matrix and $\mathsf{diag}\left(a_1,a_2,\ldots,a_n\right)$ denotes a diagonal matrix with elements $l=\{a_1,a_2,\ldots,a_n\}$. We use $(\cdot)^\dagger$ to denote the conjugate transpose, $\|\cdot\|$ is the Frobenius norm and $\tr(\cdot)$ is the trace operation. $\lambda_{\max}(\qX)$ denotes the maximum eigenvalue of a matrix $\qX$ and $\qu_{\max}(X)$ represents the eigenvector associated with $\lambda_{\max}(\qX)$. The expectation operator is denoted by $\mathcal{E}(\cdot)$ and $\mathsf{Pr}\{\cdot\}$ is probability. $K_{\nu}(z)$ is the modified Bessel function of the second kind of order $\nu$.

\section{System Model}\label{GENERAL_system_model}

We consider a basic three-node MIMO relay network consisting of one source $S$, one relay $R$, and one destination, $D$ as shown in Fig. 1. We use $N_T$ and $N_R$ to denote the number of transmit and receive antennas at $S$ and $D$, respectively. The relay is equipped with two groups of antennas; $M_R$ receive and $M_T$ transmit antennas for full-duplex operation. $S$ has no direct link to $D$, which may result from heavy path loss and high shadowing between $S$ and $D$.

\begin{figure}[t]
\centering
\includegraphics[width=0.75\linewidth]{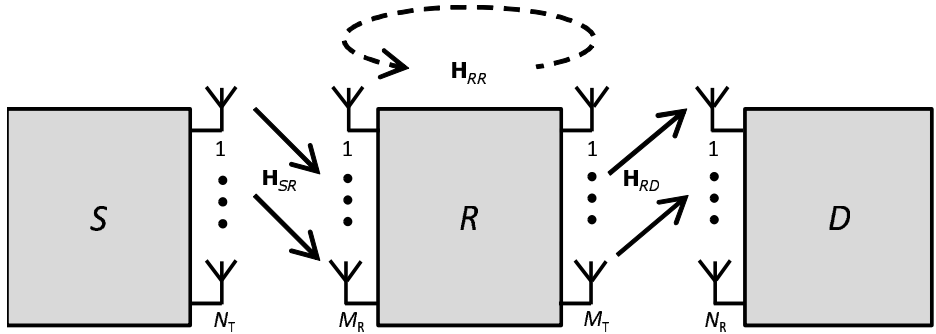}
\caption{Full-duplex MIMO relaying with multi-antenna source and destination nodes. The dashed line denotes the loopback self-interference.}
\label{FIGG1}
\end{figure}

\subsection{Channel Model}
All wireless links in the network are subject to non-selective independent Rayleigh block fading and additive white Gaussian noise (AWGN); $\qH_{SR}$ and $\qH_{RD}$ denote the $S-R$ and $R-D$ channels, respectively, while $\qH_{RR}$ denotes the LI channel. In order to reduce the effects of self-interference on system performance, an imperfect interference cancellation scheme (i.e. analog/digital cancellation) is used at $R$ and we model the residual LI channel as a fading feedback channel \cite{MIC,KRIKIDIS,RII3}. Moreover, the noise at the nodes is modeled as complex AWGN with zero mean and normalized variance. In addition, the single-input single-output channel corresponding to the $i$-th receive and the $j$-th transmit antenna from terminal $X$ to terminal $Y$, is denoted by $h^{i,j}_{XY}$ where $X \in \{S,R\}$ and $Y \in \{R,D\}$. As for the average $S-R$ and $R-D$ channel statistics; we assume $\mathcal{E}\{|h_{SR}^{i,j}|^2\}=c_{SR}$ and $\mathcal{E}\{|h_{RD}^{i,j}|^2\}=c_{RD}$. The experimental-based study in \cite{DUARTE1} has demonstrated that the amount of LI suppression achieved by an analog/digital cancellation technique is influenced by several system and hardware parameters. Since each implementation of a particular analog/digital LI cancellation scheme can be characterized by a specific residual power, a parameterization by $\qH_{RR}$ with elements satisfying $\mathcal{E}\{|h_{RR}^{i,j}|^2\}=c_{RR}$ allows these effects to be studied in a generic way \cite{RII2}. We assume that the channel coefficients remain approximately stationary for a long observation time (time slot), but change independently from one slot to another according to a Rayleigh distribution. The channel coherence time is equal to one time slot. This assumption applies to networks with a low mobility and corresponds to slow fading (block) channels where coding is performed over one block.

\subsection{System Model}
This work studies full-duplex operation at a system level using some well-known models for the characterization of the residual loop interference \cite{RII1}. We note that the developed schemes do not refer to a specific analogue or baseband implementation and can be applied to both by taking into account related practical aspects (i.e., training sequence, antenna impedance mismatch, dynamic range etc). Further implementation issues as well as more realistic radio environments (i.e., frequency selectivity) are beyond the scope of this paper.

In order to keep the complexity low, we assume that a single data stream\footnote{Single stream beamforming delivers maximum diversity/array gains and is suitable in a slow fading environment. Also, with multiple antennas at all three nodes, multiple independent data streams can be simultaneously sent, namely, multi-stream beamforming systems. In such systems, although full-duplex operation mode aimed at utilizing spectrum resources more efficiently can promise rate gains due to spatial multiplexing~\cite{RII100}, in general they experience poor error performance~\cite{ZHONG}.} is transmitted and each node employs only linear processing, i.e., $S$ applies a  precoding vector $\qt$ on the data stream, while $D$ uses a linear receive vector $\qr$ to decode the signal, where $\|\qt\|=\|\qr\|=1$. With the recent trend to increase the number of antennas at the terminals (e.g., massive MIMO), linear processing solutions offer an attractive solution for low complexity implementation. In contrast, the complexity of the optimal non linear signal detection approach grows exponentially with the number of transmit antennas~\cite{ERIKL}. The relaying operation is based on the AF policy with an amplification matrix $\qW$ that keeps the transmitted power at the relay node below the threshold $P_R$. We jointly optimize $\qt$, $\qr$ and $\qW$ to maximize the e2e system performance.

\subsubsection{Joint Precoding/Decoding Design}
ZF is chosen as the design criterion for the relay amplification matrix $\qW$, such that there is no loopback self-interference from the relay output to relay input. To simplify the problem, we further decouple $\qW$ as  $\qW = \qw_t\qw_r^\dag$, where $\qw_r$ is the receive beamforming vector and $\qw_t$ is the transmit beamforming vector both at the relay node. By fixing $\qw_r$ (or $\qw_t$), $\qw_t$ (or $\qw_r$) can be jointly optimized with $\qt$ at $S$ and $\qr$ at $R$ to realize the overall zero loopback self-interference at $R$ and maximize the e2e SNR.

\subsubsection{Antenna Selection}
AS schemes can be considered as a special case of our system model with one element of $\qr$ and $\qt$ being unity and the rest zero. Hence, only one element of $\qW$ is non-zero and this entry depends on the selected transmit and receive antennas at the relay. Specifically, in the case of AS, we assume that at each terminal, a single antenna is selected either to maximize the e2e SINR at $D$ (with optimal AS) or to maximize SNRs/SINRs associated with $S-R$, $R-R$ and $R-D$ links (with sub-optimal AS). The search complexity of the optimal scheme is high especially with a large number of antennas at each terminal, therefore, the sub-optimal schemes provide a better trade-off between implementation complexity and e2e system performance. Moreover, if $S$ transmits with a power $P_S$, we model the transmit power at $R$, as $P_S^{\alpha}$ where $0 < \alpha \leq 1$. The parameter $\alpha$ provides a dB scaling of the relay transmit power which is necessary in the presence of residual LI. Hence, $\alpha$ captures the effects of power control on the achieved performance as it allows the analysis of different relative power gains between the SINR and the SNR of the $S-R$ and $R-D$ hops, respectively. 
Although the proposed ZF precoding design operation is optimal with the use of full power at the relay ($\alpha=1)$, as we show later (in Section \ref{Outage_Probability_Analysis}), when AS schemes are implemented, an appropriate $\alpha$ can protect the MIMO relay system from error floor effects and thus a zeroth-order diversity.

\section{Joint Precoding/Decoding Design}

Based on the above system model, the equivalent $S-R$ and $R-D$ channels become
  \be\label{eqn:hsr:hrd}
  \qh_{SR}\triangleq \qH_{SR}\qt,\hspace{3mm}\mbox{and}\hspace{3mm}\qh_{RD} = \qH_{RD}^\dag\qr.
 \ee
We first assume $\qt,\qr$ are fixed and study their optimal design together with $\qw_r$ and $\qw_t^\dag$ according to different criteria.

By assuming a processing delay at $R$, given by $\tau$ \cite{RII1,RII2}, the input and the output at $R$ can be written as
  \be\label{eqn:rn}
    \qr[n] = \qh_{SR}x_S[n] + \qH_{RR} \qx_R[n] + \qn_R[n],
  \ee  \vspace{-3mm} and
  \be\label{eqn:xn}
    \qx_R[n] = \qW\qr[n-\tau],
  \ee
respectively, where $x_S[n]$ is the transmitted symbol at $S$ with zero mean, average power $P_S$ and $\qn_R$ is the $M_R\times 1$ AWGN vector with zero mean and identity covariance matrix.

Using (\ref{eqn:rn}) and (\ref{eqn:xn}) recursively, the relay output can be rewritten as
\small
    \bea\label{eqn:xn2}
    \qx_R[n] &=& \qW\qr[n-\tau] =\qW \qh_{SR}x_S[n-\tau] + \qW\qH_{RR} \qx_R[n-\tau] + \qW\qn_R[n]\notag\\
        &=&  \qW\sum_{j=0}^{\infty}\left(\qH_{RR}\qW\right)^j \left(\qh_{SR}x_S[n-j\tau-\tau] + \qn_R[n-j\tau]\right).
  \eea  \normalsize

 Note that we aim to maximize the e2e SNR, and the optimal $\qW$ should possess a minimum mean square error (MMSE) structure, which is nontrivial to solve. To simplify the signal model, and find low-complexity closed-form rather than  optimal solutions, we add the additional ZF constraint that the design of $\qW$ ensures no loopback self-interference for the full-duplex operation. To realize this, it is easy to check from (\ref{eqn:xn2}) that the following
condition is sufficient, \be
    \qW\qH_{RR}\qW=\qzero.
 \ee
 As a result, (\ref{eqn:xn2}) becomes
    \bea\label{eqn:xn3}
    \qx_R[n] &=&  \qW \left(\qh_{SR}x_S[n-\tau] + \qn_R[n]\right),
  \eea

  with the covariance matrix 
  \bea
    {\mathcal{E}}[\qx_R\qx_R^\dag] &=& P_S \qW\qh_{SR}\qh_{SR}^\dag\qW^\dag + \qW\qW^\dag.
\eea

 The relay output power is
 \be
 P_R = \tr( {\mathcal{E}}[\qx_R\qx_R^\dag])=\|\qW \qh_{SR}\|^2 P_S +  \|\qW\|^2.
 \ee

The received signal at $D$ can be written as
  \bea
    r_D[n] &=& \qh_{RD}^\dag\qx_R[n] + n_D[n]\notag\\
     &=& \qh_{RD}^\dag\qW \qh_{SR}x_S[n-\tau]  + \qh_{RD}^\dag\qW\qn_R[n] + n_D[n].
  \eea
 The e2e SINR, denoted as $\gamma$, is expressed as
\small
  \bea\label{eqn:hi1}
    \gamma &=& \frac{P_S|\qh_{RD}^\dag\qW \qh_{SR}|^2}{  \|\qh_{RD}^\dag\qW\|^2 +
     1}.
  \eea
  \normalsize
 We aim to optimize the relay processing matrices $\qW$ in order to maximize the e2e SINR. Mathematically, the optimization problem is
formulated as
 \bea\label{eqn:fd2}
    \max_{\qW} && \gamma~(\mbox{in Eq.}~\ref{eqn:hi1})\\
    \mbox{s.t.} &&  P_S\|\qW \qh_{SR}\|^2  +  \|\qW\|^2\le P_R,\notag\\
    && \qW\qH\qW=\qzero.   \notag
 \eea
 
To further simply the problem, we assume $\qW = \qw_t\qw_r^\dag$, where $\qw_r$ is the receive beamforming vector and $\qw_t$ is the transmit beamforming vector. It is noted that $\qW$ is of rank-1 and this is reasonable since there is only a single data stream. Then the ZF condition is simplified to $\qw_r^\dag\qH_{RR}\qw_t=0$. To achieve this requirement, we can design $\qw_r$ or $\qw_t$ jointly with $\qt$ and $\qr$, as described below.

\subsection{Receive ZF with $M_R>1$}
We assume maximum ratio transmission (MRT) with $\qw_t=\qh_{RD}$ and optimize $\qw_r$ based on the ZF criterion. Consequently, problem (\ref{eqn:fd2}) reduces to
 \bea\label{eqn:fdx}
    \max_{\qw_r} && \frac{P_S\|\qh_{RD}\|^4|\qw_r^\dag \qh_{SR}|^2}{  \|\qh_{RD}\|^4\|\qw_r\|^2 +
     1}\\
    \mbox{s.t.} &&  P_S\|\qh_{RD}\|^2|\qw_r^\dag\qh_{SR}|^2  +  \|\qh_{RD}\|^2\|\qw_r\|^2\le P_R,\notag\\
    && \qw_r^\dag\qH_{RR}\qh_{RD}=\qzero. \notag
 \eea
Note that the first power constraint needs to be satisfied with equality, otherwise, $\|\qw_r\|$ can be increased without violating any
constraint and this leads to a higher objective value. Hence, the objective function \eqref{eqn:fdx} can be written as
 $$ \frac{P_S\|\qh_{RD}\|^4|\qw_r^\dag \qh_{SR}|^2}{  \|\qh_{RD}\|^2(P_R - P_S\|\qh_{RD}\|^2|\qw_r^\dag\qh_{SR}|^2  ) +
     1}, $$
which is monotonically increasing in $|\qw_r^\dag \qh_{SR}|$. As a result, \eqref{eqn:fdx} is equivalent to
  \bea\label{eqn:fdx2}
    \max_{\qw_r} &&  |\qw_r^\dag \qh_{SR}|^2\\
    \mbox{s.t.} &&  P_S\|\qh_{RD}\|^2|\qw_r^\dag\qh_{SR}|^2  +  \|\qh_{RD}\|^2\|\qw_r\|^2\le P_R,\notag\\
    && \qw_r^\dag\qH_{RR}\qh_{RD}=\qzero. \notag
 \eea
Let $\qE \triangleq \qI+P_S\qh_{SR}\qh_{SR}^\dag$ and $\qE^{1/2}\qw_r=\qv_r$. With this definition, we can formulate a simple optimization problem for $\qv_r$ as follows:
\small
   \bea\label{eqn:fdx5}
    \max_{\qv_r} &&  |\qv_r^\dag\qE^{-1/2} \qh_{SR}|^2\\
    \mbox{s.t.} && \|\qv_r\|^2\le \frac{P_R}{ \|\qh_{RD}\|^2},\notag\\
    && \qv_r^\dag\qE^{-1/2}\qH_{RR}\qh_{RD}=\qzero. \notag
 \eea \normalsize
 From the ZF constraint, we know that $\qv_r$ lies in the null space of $\qE^{-1/2}\qH_{RR}\qh_{RD}$. Hence, $\qv_r = \qD\qu_r$, where $\qD \triangleq \qI - \frac{\qE^{-1/2}\qH_{RR}\qh_{RD}\qh_{RD}^\dag \qH_{RR}^\dag\qE^{-1/2}}{\|\qE^{-1/2}\qH_{RR}\qh_{RD}\|^2}$.
  The objective function in (\ref{eqn:fdx5}) then becomes $|\qu_r^\dag\qD\qE^{-1/2} \qh_{SR}|^2$ and the optimal $\qu_r$ should align with  $\qD\qE^{-1/2}
  \qh_{SR}$. Using the facts that the first power constraint should be met with equality and $\qD$ is idempotent, 
   we can express the optimal solutions of (\ref{eqn:fdx5}) and (\ref{eqn:fdx2}) as \small
     \begin{align}\label{eqn:wv}
     \qv_r&=\frac{\qD\qE^{-1/2} \qh_{SR}}{\|\qD\qE^{-1/2} \qh_{SR}\|}\sqrt{\frac{P_R}{ \|\qh_{RD}\|^2}},~~\mbox{and}\\\nonumber
     \qw_r&=\frac{\qE^{-1/2} \qD\qE^{-1/2} \qh_{SR}}{\|\qD\qE^{-1/2} \qh_{SR}\|}\sqrt{\frac{P_R}{ \|\qh_{RD}\|^2}}.
     \end{align} \normalsize
The objective value in (\ref{eqn:fdx}) involves $|\qw_r^\dag\qh_{SR}|^2$  and $\|\qw_r\|^2$ which, from \eqref{eqn:wv}, are given by
\small
 \bea \label{eqn:ar1}
    |\qw_r^\dag\qh_{SR}|^2 &= &\left|\frac{\qh_{SR}^\dag\qE^{-1/2} \qD\qE^{-1/2} \qh_{SR}}{\|\qD\qE^{-1/2} \qh_{SR}\|}\right|^2{\frac{P_R}{
    \|\qh_{RD}\|^2}}\notag\\
    &= &  {\frac{P_R}{ \|\qh_{RD}\|^2}}\qh_{SR}^\dag\qE^{-1/2} \qD\qE^{-1/2} \qh_{SR}\notag\\
 &= &  {\frac{P_R}{ \|\qh_{RD}\|^2}} \left(\|\qh_{SR}^\dag\qE^{-1/2}\|^2 - \frac{\|\qh_{SR}^\dag\qE^{-1}\qH_{RR}\qh_{RD}\|^2}{\|\qE^{-1/2}\qH_{RR}\qh_{RD}\|^2}\right)\notag\\
        &= &    {\frac{P_R}{ \|\qh_{RD}\|^2 }} \frac{\|\qH_{RR}\qh_{RD}\|^2\|\qh_{SR}\|^2
      -   |\qh_{SR}^\dag\qH_{RR}\qh_{RD}|^2}
  {  \|\qH_{RR}\qh_{RD}\|^2  +P_S (\|\qH_{RR}\qh_{RD}\|^2\|\qh_{SR}\|^2 - |\qh_{SR}^\dag\qH_{RR}\qh_{RD}|^2) },
 \eea \normalsize
and \small
\bea
    \|\qw_r\|^2 &=& \frac{P_R}{\|\qh_{RD}\|^2} - P_S|\qw_r^\dag\qh_{SR}|^2 \\
    &=& \frac{P_R}{\|\qh_{RD}\|^2} - {\frac{P_SP_R}{ \|\qh_{RD}\|^2 }}    \frac{  \|\qH_{RR}\qh_{RD}\|^2\|\qh_{SR}\|^2
      -   |\qh_{SR}^\dag\qH_{RR}\qh_{RD}|^2}
  {  \|\qH_{RR}\qh_{RD}\|^2  +P_S (\|\qH_{RR}\qh_{RD}\|^2\|\qh_{SR}\|^2  -  |\qh_{SR}^\dag\qH_{RR}\qh_{RD}|^2) } \\
  &=& \frac{P_R}{\|\qh_{RD}\|^2}   \frac{ \|\qH_{RR}\qh_{RD}\|^2 }
  {  \|\qH_{RR}\qh_{RD}\|^2  +P_S (\|\qH_{RR}\qh_{RD}\|^2\|\qh_{SR}\|^2  -  |\qh_{SR}^\dag\qH_{RR}\qh_{RD}|^2) }.
\eea \normalsize
Using \eqref{eqn:ar1} and (17) in \eqref{eqn:fdx}, the achievable e2e SNR can be derived as
 \bea\label{eqn:SNR:R}
    \gamma
          &= & \frac{P_S \|\widehat\qD\qh_{SR}\|^2 P_R \|\qh_{RD}\|^2}
    {P_S \|\widehat\qD\qh_{SR}\|^2 + P_R\|\qh_{RD}\|^2 + 1},
  \eea
where $\widehat\qD\triangleq  \qI - \frac{\qH_{RR}\qh_{RD}\qh_{RD}^\dag \qH_{RR}^\dag}{\|\qH_{RR}\qh_{RD}\|^2}$.

Next, we can address the design of $\qt$ and $\qr$. Notice from (\ref{eqn:hsr:hrd}) that $\qt$ and $\qr$ are embedded in $
\|\widehat\qD\qh_{SR}\|^2$ and  $\|\qh_{RD}\|^2$, respectively, so we propose the following solution to separately optimize $\qt$ and $\qr$:
\begin{align}
\label{smith2}
    \qt^* &= \arg \max_{\|\qt\|=1} \|\widehat\qD\qh_{SR}\|^2 =\arg \max_{\|\qt\|=1} \|\widehat\qD\qH_{SR}\qt\|^2\\\nonumber
    &= \qu_{\max}(\qH_{SR}^\dag\widehat\qD\qH_{SR}),
\end{align}
and
\begin{align}
\label{smith1}
    \qr^* &= \arg \max_{\|\qr\|=1} \|\qh_{RD}\|^2  =\arg \max_{\|\qr\|=1} \|\qH^\dag_{RD}\qr\|^2\\\nonumber
    &= \qu_{\max}(\qH_{RD}^\dag\qH_{RD}),
\end{align}
 where we have used the fact that $\widehat\qD$ is idempotent. Note that $\widehat\qD$ also depends on $\qr$ via $\qh_{RD}$, so the above solutions may not be optimal. Nevertheless, the choice of $\qr^*$ in \eqref{smith1} uniquely maximizes $\|\qh_{RD}\|^2$ and given this choice of $\qr$, $\qt^*$ in \eqref{smith2} uniquely maximizes $\|\widehat\qD\qh_{SR}\|^2$. Hence, this approach is very appealing and these simple closed-form solutions facilitate both the precoder/receive vector design and performance analysis.

 Substituting $\qt^*$ and $\qr^*$ back into \eqref{eqn:SNR:R}, the e2e SNR can be
expressed as
 \bea\label{eqn:SNR:R2}
    \gamma
          &= & \frac{P_S \|\widehat\qD\qH_{SR}\|_2^2 P_R \|\qH_{RD}\|_2^2 }
    {P_S \|\widehat\qD\qH_{SR}\|_2^2+P_R\|\qH_{RD}\|_2^2 +1},
  \eea
where $\|\qX\|_2^2 = \lambda_{\max}(\qX\qX^\dag)$.

\subsection{Transmit ZF with $M_T>1$}
We assume   that $\qw_r=\qh_{SR}$, i.e., the relay employs a maximal-ratio combining (MRC) receive beamforming vector, and optimizes the transmit ZF vector $\qw_t$. In this case, we can  simplify problem (\ref{eqn:fd2}) as:
 \small
 \bea\label{eqn:fd3}
    \max_{\qw_t} && \frac{P_S|\qh_{RD}^\dag\qw_t|^2 \|\qh_{SR}\|^4}{
    |\qh_{RD}^\dag\qw_t|^2\|\qh_{SR}\|^2 + 1}\\
    \mbox{s.t.} &&   \|\qw_t\|^2\le \frac{P_R}{\|\qh_{SR}\|^4 P_S +  \|\qh_{SR}\|^2 },\notag\\
    && \qh_{SR}^\dag\qH_{RR}\qw_t=0,\notag
 \eea \normalsize
 or equivalently using monotonicity,
 \small
  \bea\label{eqn:fd4}
    \max_{\qw_t} &&  |\qh_{RD}^\dag\qw_t|^2\\
    \mbox{s.t.} &&   \|\qw_t\|^2\le \frac{P_R}{\|\qh_{SR}\|^4 P_S +  \|\qh_{SR}\|^2 },\notag\\
    && \qh_{SR}^\dag\qH_{RR}\qw_t=0.\notag
 \eea \normalsize
Following the same procedure employed to obtain (\ref{eqn:wv}), the solution of \eqref{eqn:fd4} is given by
 \be
     \qw_t^* =  {\sqrt{\frac{P_R}{\|\qh_{SR}\|^4 P_S +  \|\qh_{SR}\|^2
     }}}\frac{\qB\qh_{RD}}{{\|\qB\qh_{RD}\|}},
 \ee
where we have defined $\qB\triangleq \qI -\frac{\qH_{RR}^\dag\qh_{SR}\qh_{SR}^\dag\qH_{RR}}{\|\qh_{SR}^\dag\qH_{RR}\|^2}$. With $\qw_t^*$, the optimized e2e SNR can be expressed as
\bea\label{eqn:SNR:t}
    \gamma = \frac{  P_S \|\qh_{SR}\|^2 P_R\|\qB\qh_{RD}\|^2}{
        {P_S\|\qh_{SR}\|^2  + {P_R}\|\qB\qh_{RD}\|^2+ 1 }}.
\eea Similar to the receive ZF scheme, we propose the following solutions for $\qt$ and $\qr$ (which may not be optimal)
\begin{align}
    \qt^*& = \arg \max_{\|\qt\|=1} \|\qh_{SR}\|^2  =\arg \max_{\|\qt\|=1} \|\qH_{SR}\qt\|^2\\\nonumber
    & = \qu_{\max}(\qH_{SR}^\dag \qH_{SR}),
\end{align}
and
\begin{align}
  \qr^*& = \arg \max_{\|\qr\|=1} \|\qB\qh_{RD}\|^2  =\arg \max_{\|\qr\|=1} \|\qB\qH_{RD}\qr\|^2\\\nonumber
  & = \qu_{\max}(\qH_{RD}^\dag\qB\qH_{RD}),
\end{align}
respectively. Finally, substituting $\qt^*$ and $\qr^*$ into \eqref{eqn:SNR:t}, the e2e SNR can be expressed as
\bea\label{eqn:SNR:t2}
    \gamma = \frac{  P_S \|\qH_{SR}\|_2^2 {P_R}\|\qB\qH_{RD}^\dag\|_2^2}{
         P_S {\|\qH_{SR}\|_2^2 + {P_R}\|\qB\qH_{RD}^\dag\|_2^2 + 1 }}.
\eea

\section{Antenna Selection}
This section deals with the problem of AS for the full-duplex MIMO relay channel considered. AS is proposed as an alternative to the e2e optimization and is particularly relevant to systems with stricter computational/energy constraints. Full-duplex relay AS introduces new design challenges due to the presence of LI and differs from the existing  body of AS literature in several ways. As explained below, with full-duplex operation, several AS choices that provide different performance/complexity tradeoff exist while a straightforward AS strategy (see for e.g. \cite{HWCNC10}) can be used to maximize the performance in half-duplex AS systems. Moreover, power allocation is an important issue with different full duplex AS schemes while half-duplex AS schemes can use full power at the relay (in the absence of LI).

The AF process at $R$ employs the conventional amplification factor \cite[Eq. (4)]{RII2} which guarantees the stability of the relay and prevents oscillation. This particular choice of amplification process is also simple to use since $R$ can adaptively adjust its transmit power to a constant level. In this case, the instantaneous e2e SINR is expressed as~\cite{RII2,RII3}
\begin{align}
\label{SINR_GEN_AS}
\gamma^{i,j,k,l} & = \frac{\frac{\gamma^{i,j}_{SR}}{\gamma^{i,l}_{RR}+1}\gamma^{k,l}_{RD}}{\frac{\gamma^{i,j}_{SR}}{\gamma^{i,l}_{RR}+1}+\gamma^{k,l}_{RD}+1},
\end{align}
where $\gamma^{i,j}_{SR} = P_{S}|h^{i,j}_{SR}|^2$, and $\gamma^{k,l}_{RD} = P_{S}^{\alpha}|h^{k,l}_{RD}|^2$ are the instantaneous SNRs of the $S-R$ and the $R-D$ links while $\gamma^{i,l}_{RR} = P_{S}^{\alpha}|h^{i,l}_{RR}|^2$ is the instantaneous interference-to-noise ratio (INR) of the $R-R$ link. In order to facilitate the analysis of the outage probability in Section \ref{OUT_AS_Analysis}, we also restate the average SNRs of the $S-R$ and the $R-D$ links as $\bar{\gamma}_{SR} \triangleq P_{S} {c}_{SR}$ and $\bar{\gamma}_{RD} \triangleq P_{S}^{\alpha}{c}_{RD}$, respectively. Moreover, $\bar{\gamma}_{RR} \triangleq P_{S}^{\alpha} {c}_{RR}$ is the average INR of the $R-R$ link.

\subsection{Optimal Antenna Selection}
Denote the selected receive and transmit antenna indexes at $R$ and $S$, and the receive and transmit antenna indexes at $D$ and $R$ are by $I, J, K, L$, respectively. The optimal AS (OP AS) scheme can be expressed as
\begin{align}
\{I, J, K, L \} = \mathop{\text{argmax}}\limits_{\substack{1 \leq i \leq M_{R}, 1 \leq j \leq N_{T}\\ 1 \leq k \leq N_{R}, 1 \leq l \leq M_{T}}} \left(\gamma^{i,j,k,l}\right).
\end{align}
The OP AS scheme maximizes the e2e SINR, however it has a high computation and implementation complexity. In a centralized
architecture, a central unit requires the knowledge of all links ($S-R$, $R-R$ and $R-D$) in order to decide on the selected antennas.

\subsection{$\max$-$\max$ Antenna Selection}
The $\max-\max$ AS (MM AS) scheme selects the best $S-R$ and $R-D$ links without considering the LI and can be expressed as
\begin{align}
\label{mmselect}
\{I,J\} = \mathop{\text{argmax}}\limits_{1 \leq i \leq M_{R},1 \leq j \leq N_{T}}\left(\gamma^{i,j}_{SR}\right), \hspace{3mm} \{K,L\} &= \mathop{\text{argmax}}\limits_{1 \leq k \leq N_{R},1 \leq l \leq M_{T}}\left(\gamma^{k,l}_{RD}\right).
\end{align}
Note that the MM AS scheme, which is SNR optimal in  conventional half-duplex relaying \cite{HWCNC10}, becomes strictly sub-optimal in full-duplex relaying since it does not take into account the effect of LI. However, the MM AS scheme can be easily implemented by estimating the $S-R$ channels at $R$ and using channel feedback (on the $R-D$ link) from $D$ to $R$, related to the selected antenna index $K$.

\subsection{Partial Antenna Selection}
The partial AS (PR AS) scheme\footnote{The name for this AS scheme was adopted in the same spirit where selection schemes based on the first-hop CSI are identified as partial relay selection in the literature~\cite{KRIKIDIS}.} simplifies the selection problem by decoupling the two relaying hops according to the following rule
\begin{align}
\{I,J,L\}  = \mathop{\text{argmax}}\limits_{1 \leq i \leq M_{R}, 1 \leq j \leq N_{T}, 1 \leq l \leq M_{T}}\left(\frac{\gamma^{i,j}_{SR}}{\gamma^{i,l}_{RR}+1}\right), \hspace{3mm} \{K\}  = \mathop{\text{argmax}}\limits_{1 \leq k \leq N_{R}} \left(\gamma^{k,L}_{RD}\right).
\end{align}
The PR AS scheme provides a good performance/implementation complexity trade-off since it reduces the searching set of the optimal solution while it also takes into account the LI. It is worth noting that channel feedback from $D$ to $R$ is not required since the relay transmit antenna is selected independently of the second hop.

\subsection{Loop Interference Antenna Selection}
The loop interference AS (LI AS) scheme selects the receive/transmit antennas in order to minimize the effects of LI according to
\begin{align}
\{I,L\}  = \mathop{\text{argmin}}\limits_{1 \leq i \leq M_{R}, 1 \leq l \leq M_{T}}\left(\gamma^{i,l}_{RR}\right), \hspace{3mm} \{J\}  = \mathop{\text{argmax}}\limits_{1 \leq j \leq N_{T}} \left(\gamma^{I,j}_{SR}\right), \hspace{3mm} \{K\} = \mathop{\text{argmax}}\limits_{1 \leq k \leq N_{R}} \left(\gamma^{k,L}_{RD}\right).
\end{align}
This scheme is analogous to the LI suppression policies with relay precoders proposed in \cite{RII1,SUNG}. The LI AS aims to minimize the deleterious effects of LI, while some improvement in the $S-R$, $R-D$ channels is also extracted by selecting antennas at $S$ and $D$.

\section{Outage Probability Analysis}\label{Outage_Probability_Analysis}
In this section, we study the outage probability of the precoding/decoding designs as well as the AS schemes presented in Sections III and IV, respectively. We derive exact expressions for the outage probability and based on these results, the asymptotic behavior is also studied to reveal important insights such as the diversity order.

\subsection{Joint Precoding/Decoding Designs}
The rate outage probability, $P_{\mathsf{out}}$, is defined as the probability that the instantaneous mutual information, $\mathcal{I}=\log_2\left(1+\gamma\right)$, falls below a target rate of $R_0$ bits per channel use (BPCU). Hence,
\begin{align}
P_{\mathsf{out}}&=\mathsf{Pr}\left(\log_2\left(1+\gamma\right) \leq R_0\right) = F_{\gamma}\left(\gamma_T\right),
\end{align}
where $\gamma_T = 2^{R_0}-1$ and $F_{\gamma}\left(\cdot\right)$ is the cumulative distribution function (cdf) of the e2e SNR.
\subsubsection{Receive ZF}
From (\ref{eqn:SNR:R2}), we can now derive the outage probability of the system. To this end, we first note that $\|\widehat\qD\qH_{SR}\|_2^2=\lambda_{\max}\left( \qH_{SR}^\dagger \widehat\qD^\dagger \widehat\qD \qH_{SR}\right)$ can be written as
\small
\begin{align}
\label{eqn:pre_analysis_e1}
\|\widehat\qD\qH_{SR}\|_2^2 & = \lambda_{\max}\left( \qH_{SR}^\dagger \left(\qI - \frac{\qH_{RR}\qh_{RD}\qh_{RD}^\dag \qH_{RR}^\dag}{\|\qH_{RR}\qh_{RD}\|^2}\right) \qH_{SR}\right)\nonumber\\
& = \lambda_{\max}\left( \qH_{SR}^\dagger \mathbf{\Phi}^\dagger \left(\qI-\mathsf{diag}\left(1,0,\ldots,0\right)\right) \mathbf{\Phi}\qH_{SR}\right)\nonumber\\
& = \lambda_{\max}\left( \widehat{\qH}_{SR}^\dagger \mathsf{diag}\left(0,1,\ldots,1\right) \widehat{\qH}_{SR}\right)\nonumber\\
& = \lambda_{\max}\left(\breve{\qH}_{SR}^\dagger \breve{\qH}_{SR}\right),
\end{align} \normalsize
where $\mathbf{\Phi}$ is a unitary matrix, $\widehat{\qH}_{SR}=\mathbf{\Phi}\qH_{SR}$ and $\breve{\qH}_{SR}$ is a $(M_{R}-1) \times N_T$ matrix. In \eqref{eqn:pre_analysis_e1}, the first equality follows from the fact that $\widehat{\qD} = \widehat\qD^\dagger \widehat\qD $. The second equality is due to the eigen decomposition ($\frac{\qH_{RR}\qh_{RD}}{\|\qH_{RR}\qh_{RD}\|}$ is a $M_R \times 1$ normalized column vector and has rank 1). Hence, $\|\widehat\qD\qH_{SR}\|_2^2$ is the maximum eigenvalue of a Wishart matrix $\left(\breve{\qH}_{SR}^\dagger \breve{\qH}_{SR}\right)$ with dimensions $(M_R-1) \times N_T$.

We now derive the exact outage probability with receive ZF using the result for $\|\widehat\qD\qH_{SR}\|_2^2$ in \eqref{eqn:SNR:R2} in conjunction with $\|\qH_{RD}\|_2^2$. The required cdf of the e2e SNR can be derived by adopting a similar approach as in \cite[Appendix I]{GAYAN2}. Specifically, we can express the cdf of $\gamma$ as $F_{\gamma}(\gamma_T)=\mathsf{Pr}\left(\frac{\gamma_{SR}\gamma_{RD}}{\gamma_{SR}+\gamma_{RD}+1} < \gamma_T\right) = 1-\int^{\infty}_{0} \bar{F}_{\gamma_{RD}}\left(\frac{(\gamma_T+y+1)\gamma_T}{y}\right)f_{\gamma_{SR}}(\gamma_T+y)dy$, where $\bar{F}_{\gamma_{RD}}(x)$ is the complementary cdf of $\gamma_{RD}$, and $f_{\gamma_{SR}}(x)$ is the probability density function  (pdf) of $\gamma_{SR}$, with $\gamma_{SR}=P_S \|\widehat\qD\qH_{SR}\|_2^2$ and $\gamma_{RD}=P_R \|\qH_{SR}\|_2^2$. By using \cite[Eq. (23)]{Dighe}, we can obtain the pdf of $\gamma_{SR}$ and the cdf of $\gamma_{RD}$ as
\begin{align*}
f_{\gamma_{SR}}(x)=\sum^{\min(N_T,M_R-1)}_{a=1}\sum^{(N_T+M_R-1)a-2a^2}_{b=|N_T-M_R+1|}\frac{a^{b+1}d_1(a,b)}{(b)!\bar{\gamma}_{SR}^{b+1}}x^b e^{-\frac{ax}{\bar{\gamma}_{SR}}},
\end{align*}
and 
\begin{align*}
F_{\gamma_{RD}}(x)=1-\sum^{\min(M_T,N_R)}_{k=1}\sum^{(M_T+N_R)k-2k^2}_{l=|M_T-N_R|}\sum^{l}_{m=0}\frac{k^{m}d_2(k,l)}{(m)!\bar{\gamma}_{RD}^{m}}x^m e^{-\frac{kx}{\bar{\gamma}_{RD}}}, 
\end{align*}
respectively, where the average SNR of the $S-R$ and $R-D$ links are given by $\bar{\gamma}_{SR}=P_S{c}_{SR}$ and $\bar{\gamma}_{RD}=P_R{c}_{RD}$. The coefficients, $d_l(i,j)$, $l=1,2$ are given in \cite{Dighe} for some system configurations and can be efficiently computed using the algorithm in \cite{Maaref}. We now substitute the above pdf and cdf into the integral representation of $F_{\gamma}(\gamma_T)$ and solve it in closed-form using \cite[Eq. (3.471.9)]{M21} to yield
\begin{align}
F_{\gamma}(\gamma_T)&=1-\sum_{a=1}^{s_1}\sum_{b=|N_{T}-M_{R}+1|}^{(N_{T}+M_{R}-1)a-2a^2}\sum_{k=1}^{s_2}\sum_{l=|M_{T}-N_{R}|}^{(M_{T}+N_{R})k-2k^2}\sum^{l}_{m=0}\\\nonumber &\times \sum^{m}_{u=0} \sum^{b}_{v=0}\frac{2\binom{m}{u}\binom{b}{v}d_1(a,b)d_2(k,l)k^{\frac{u+v+m+1}{2}}\gamma_T^{
\frac{m+2b+u-v+1}{2}}\left(1+\gamma_T\right)^{\frac{m-u+v+1}{2}}}{b!m!a^{\frac{u+v-m-2b-1}{2}}\bar{\gamma}_{SR}^{\frac{2b-u-v+m+1}{2}}\bar{\gamma}_{RD}^{\frac{u+v+m+1}{2}}}\\\nonumber
&\times e^{-\left(\frac{a}{\bar{\gamma}_{SR}}+\frac{k}{\bar{\gamma}_{RD}}\right)\gamma_T} K_{u+v-m+1}\left(2\sqrt{\frac{ak\left(1+\gamma_T\right)\gamma_T}{\bar{\gamma}_{SR}\bar{\gamma}_{RD}}}\right),
\end{align}
where $s_1 = \min\left(N_{T},M_{R}-1\right)$ and $s_2 = \min\left(M_{T},N_{R}\right)$. 

In order to further obtain insights, such as diversity order, we now present a simplified asymptotic outage probability. Specifically, we adopt the upper bound, $\gamma \leq \min\left(\gamma_{SR},\gamma_{RD}\right)$, to $\gamma$. This bound is tight for medium-to-high SNR values and in \cite{FANG} it was shown that it is also asymptotically-exact in the high SNR regime \cite{FANG}. Therefore, using simple order statistics we can express the asymptotic cdf of the e2e SNR as  $F_{\gamma}^{\infty}(x) = F_{\gamma_{SR}^\infty}(x)+F_{\gamma_{RD}^\infty}(x)-F_{\gamma_{SR}^\infty}(x)F_{\gamma_{RD}^\infty}(x)$.

It can be easily shown that at high SNRs, $F_{\gamma}^{\infty}(x)$ can be approximated by a single term polynomial approximation. To see this, we first need polynomial approximations for $\gamma_{SR}$ and $\gamma_{RD}$. These results can be borrowed from \cite[Eq. (7)]{Zhou} and with the aid of $F_{\gamma}^{\infty}(x)$ we can show that
\begin{align}
\label{HSNR_RZF}
P^{\infty}_{\mathsf{out}}= \left\{ \begin{array}{ll}
         \frac{\prod_{k=0}^{s_1-1}k!}{\prod_{k=0}^{s_1-1}(t_1+k)!}\left(\frac{\gamma_T}{\bar{\gamma}_{SR}}\right)^{N_{T}(M_{R}-1)} & \mbox{$N_{T}(M_{R}-1)< M_{T}N_{R}$},\\ \frac{\prod_{k=0}^{s_1-1}k!}{\prod_{k=0}^{s_1-1}(t_1+k)!}\left(\frac{\gamma_T}{\bar{\gamma}_{SR}}\right)^{N_{E}} + \frac{\prod_{k=0}^{s_2-1}k!}{\prod_{k=0}^{s_2-1}(t_2+k)!}\left(\frac{\gamma_T}{\bar{\gamma}_{RD}}\right)^{N_{E}} & \mbox{$N_{T}(M_{R}-1)= M_{T}N_{R}=N_{E}$},\\
         \frac{\prod_{k=0}^{s_2-1}k!}{\prod_{k=0}^{s_2-1}(t_2+k)!}\left(\frac{\gamma_T}{\bar{\gamma}_{RD}}\right)^{M_{T}N_{R}} & \mbox{$N_{T}(M_{R}-1)> M_{T}N_{R}$},\end{array} \right.
\end{align}
where $t_1 = \max\left(N_{T},M_{R}-1\right)$ and $t_2 = \max\left(M_{T},N_{R}\right)$. By inspecting \eqref{HSNR_RZF}, we see that our full-duplex receive ZF design achieves a diversity order of $\min\left(N_{T}(M_{R}-1), M_{T}N_{R} \right)$.

\subsubsection{Transmit ZF}
Using an equivalent approach to that used for the receive ZF scheme and omitting details for conciseness, the exact outage probability can be expressed as
\begin{align}
F_{\gamma}(\gamma_T)&=1-\sum_{a=1}^{s_3}\sum_{b=|N_{T}-M_{R}|}^{(N_{T}+M_{R})a-2a^2}\sum_{k=1}^{s_4}\sum_{l=|M_{T}-N_{R}-1|}^{(M_{T}+N_{R}-1)k-2k^2}\sum^{l}_{m=0}\\\nonumber &\times \sum^{m}_{u=0} \sum^{b}_{v=0}\frac{2\binom{m}{u}\binom{b}{v}d_1(a,b)d_2(k,l)k^{\frac{u+v+m+1}{2}}\gamma_T^{
\frac{m+2b+u-v+1}{2}}\left(1+\gamma_T\right)^{\frac{m-u+v+1}{2}}}{b!m!a^{\frac{u+v-m-2b-1}{2}}\bar{\gamma}_{SR}^{\frac{2b-u-v+m+1}{2}}\bar{\gamma}_{RD}^{\frac{u+v+m+1}{2}}}\\\nonumber
&\times e^{-\left(\frac{a}{\bar{\gamma}_{SR}}+\frac{k}{\bar{\gamma}_{RD}}\right)\gamma_T} K_{u+v-m+1}\left(2\sqrt{\frac{ak\left(1+\gamma_T\right)\gamma_T}{\bar{\gamma}_{SR}\bar{\gamma}_{RD}}}\right),
\end{align}
where $s_3 = \min\left(N_{T},M_{R}\right)$ and $s_4 = \min\left(M_{T}-1,N_{R}\right)$.

Furthermore, we can express the asymptotic outage probability of transmit ZF as
\begin{align}
\label{HSNR_TZF}
P^{\infty}_{\mathsf{out}}= \left\{ \begin{array}{ll}
         \frac{\prod_{k=0}^{s_3-1}k!}{\prod_{k=0}^{s_3-1}(t_3+k)!}\left(\frac{\gamma_T}{\bar{\gamma}_{SR}}\right)^{N_{T}M_{R}} & \mbox{$N_{T}M_{R}< (M_{T}-1)N_{R}$},\\ \frac{\prod_{k=0}^{s_3-1}k!}{\prod_{k=0}^{s_3-1}(t_3+k)!}\left(\frac{\gamma_T}{\bar{\gamma}_{SR}}\right)^{M_{E}} + \frac{\prod_{k=0}^{s_4-1}k!}{\prod_{k=0}^{s_4-1}(t_4+k)!}\left(\frac{\gamma_T}{\bar{\gamma}_{RD}}\right)^{M_{E}} & \mbox{$N_{T}M_{R}= (M_{T}-1)N_{R}=M_{E}$},\\
         \frac{\prod_{k=0}^{s_4-1}k!}{\prod_{k=0}^{s_4-1}(t_4+k)!}\left(\frac{\gamma_T}{\bar{\gamma}_{RD}}\right)^{(M_{T}-1)N_{R}} & \mbox{$N_{T}M_{R}> (M_{T}-1)N_{R}$},\end{array} \right.
\end{align}
where $t_3 = \max\left(N_{T},M_{R}\right)$ and $t_4 = \max\left(M_{T}-1,N_{R}\right)$. From Eq. \eqref{HSNR_TZF} we see that with transmit ZF, a diversity order of $\min\left(N_{T}M_{R}, (M_{T}-1)N_{R} \right)$ can be achieved.

On the other hand, half-duplex MIMO hop-by-hop (MRT/MRC) beamforming exhibits a diversity order of $\min\left(N_{T}M_{R}, M_{T}N_{R} \right)$. As a result, although half-duplex hop-by-hop beamforming delivers a superior diversity performance in general, in certain antenna configurations, half-duplex hop-by-hop beamforming and full-duplex ZF designs offer the same diversity.

\subsection{Antenna Selection}\label{OUT_AS_Analysis}
In this subsection, we investigate the outage probability of the proposed full-duplex based AS schemes. We derive exact as well as approximate outage expressions when $P_S \rightarrow \infty$ for comparison of the proposed AS schemes. By considering the definition of the outage probability, we can write\footnote{In the following subsections, the statistical distributions of $\gamma^{I,J}_{SR}$, $\gamma^{I,L}_{RR}$ and $\gamma^{K,L}_{RD}$ may differ depending on the AS scheme. Any remark concerning the distributions of these RVs is strictly limited to the particular AS scheme.}
\begin{align}
\label{eqn:star}
P_{\star}=\mathsf{Pr}\left\{\log_2\left(1+\frac{\frac{\gamma^{I,J}_{SR}}{\gamma^{I,L}_{RR}+1}\gamma^{K,L}_{RD}}{\frac{\gamma^{I,J}_{SR}}{\gamma^{I,L}_{RR}+1}+\gamma^{K,L}_{RD}+1}\right)<R_0\right\}.
\end{align}
For ``optimal'', ``$\max-\max$'', ``partial'' and ``loop interference'' AS schemes, the subscript $\star$ in \eqref{eqn:star} refers to $\text{OP}$, $\text{MM}$, $\text{PR}$ and $\text{LI}$, respectively.

\subsubsection{Optimal Antenna Selection}
Let $\gamma_{\text{OP}}$ denote the e2e SINR at $D$ for the OP AS scheme. The outage probability of the OP AS scheme can be written as
\begin{align}
P_{\text{OP}} = F_{\gamma_{\text{OP}}}\left(\gamma_T\right),
\end{align}
where $F_X(\cdot)$ denotes the cdf of the random variable (RV), $X$. Obtaining an analytical expression for $P_{\text{OP}}$ appears to be a cumbersome problem due to the dependencies between the SINR variables being maximized. Therefore, we have performed simulations to evaluate the outage performance of the OP AS scheme in Section V. Further, under some special antenna configurations, for example with $M_{\text{R}}=M_{\text{T}}=1$, the OP AS scheme is equivalent to the MM AS scheme for which an analytical expression is presented below.

We now state the asymptotic behavior of the OP AS scheme in \emph{Proposition 1}.

\emph{Proposition 1:} The outage probability of the OP AS scheme as $P_S \rightarrow \infty$ can be approximated by
\begin{align}
\label{OSHSNR}
P_{\text{OP}} \approx \mathcal{C}_1\left(\frac{\bar{\gamma}_{RR}\gamma_T}{\bar{\gamma}_{SR}}\right)^{N_{T}M_{R}}+\mathcal{C}_2 \left(\frac{\gamma_T}{\bar{\gamma}_{RD}}\right)^{M_{T}N_{R}},
\end{align}
where $\mathcal{C}_1>0$ and $\mathcal{C}_2>0$ are two positive constants.

\begin{proof}
We first lower bound $\gamma_{\text{OP}}$ by $\gamma_{\text{MM}}$, where $\gamma_{\text{MM}}$ is the SINR of the suboptimal MM AS scheme. In the following subsection, we show that as $P_S$ tends to infinity, the corresponding upper bound, $P_{\text{OP}} \leq P_{\text{MM}}$, can be approximated by $P_{\text{MM}} \approx \left(N_{T}M_{R}\right)!\left(\frac{\bar{\gamma}_{RR}\gamma_T}{\bar{\gamma}_{SR}}\right)^{N_{T}M_{R}}+\left(\frac{\gamma_T}{\bar{\gamma}_{RD}}\right)^{M_{T}N_{R}}$. Next we upper bound $\gamma_{\text{OP}}$ by $\gamma_{\text{UB}}$,\footnote{The SINR upper bound, $\gamma_{\text{UB}}$ corresponds to a ``virtual'' system in which transmit/receive AS is decoupled to consider the best $S-R$ and $R-D$ links and the weakest LI ($R-R$) link, respectively, since such a strategy will maximize the e2e SINR in \eqref{SINR_GEN_AS}. However, clearly such a AS scheme is not possible in our system, since selecting a particular transmit/receive antenna pair at $R$ will automatically fix the LI link, i.e., AS for the links can not be performed independently.} defined as $\gamma_{\text{UB}} \triangleq \frac{\frac{X_1}{Y} X_2}{\frac{X_1}{Y} + X_2 +1}$ where $X_1$ and $X_2$ are the maximum of $N_{T}M_{R}$ and $M_{T}N_{R}$ exponential RVs with parameters, $\bar{\gamma}_{SR}$ and $\bar{\gamma}_{RD}$, respectively, while $Y$ is a RV chosen as the minimum of $M_{T}M_{R}$ exponential RVs with parameter $\bar{\gamma}_{RR}$. As $P_S$ tends to infinity, we can show that the corresponding lower bound, $P_{\text{OP}} > P_{\text{LB}}$ can be approximated by $P_{\text{LB}} \approx \frac{(N_{T}M_{R})!}{\left(M_{T}M_{R}\right)^{N_{T}M_{R}}}\left(\frac{\bar{\gamma}_{RR}\gamma_T}{\bar{\gamma}_{SR}}\right)^{N_{T}M_{R}}+\left(\frac{\gamma_T}{\bar{\gamma}_{RD}}\right)^{M_{T}N_{R}}$. Since the upper and lower bounds of $P_{\text{OP}}$ have the same diversity order, \eqref{OSHSNR} follows and the proof is completed.
\end{proof}

Using the above asymptotic result, we now derive the optimal $\alpha$ to yield the power allocation solution at the relay. Following the respective definitions and expressing $\bar{\gamma}_{SR}$, $\bar{\gamma}_{RR}$ and $\bar{\gamma}_{RD}$ explicitly in terms of $P_S$, we see that the first term in \eqref{OSHSNR} decays as ${P_S^{-(1-\alpha)N_{T}M_{R}}}$ while the second term decays as $P_S^{-\alpha M_{T}N_{R}}$. Therefore, depending on the value of $\alpha$, the first or the second term in \eqref{OSHSNR} becomes dominant and determines the total asymptotic outage probability. Outage minimization from a diversity perspective occurs when $(1-\alpha)N_{T}M_{R}=\alpha M_{T}N_{R}$ and we have
\begin{align}
\alpha^{\text{OP}}_{\mathsf{opt}} = \frac{N_{T}M_{R}}{N_{T}M_{R}+ M_{T}N_{R}},
\end{align}
with $P^{\alpha^{\text{OS}}_{\mathsf{opt}}}_{S}$ as the optimal power allocation solution at the relay. Moreover, the highest diversity order, $d_{\text{max,OP}}$, achieved with the OP AS scheme is given by
\begin{align}
d_{\text{max,OP}}=\frac{1}{\left(M_{T}N_{R}\right)^{-1}+\left(N_{T}M_{R}\right)^{-1}}.
\end{align}

\subsubsection{$\max-\max$ Antenna Selection}
With this scheme, $\gamma^{I,J}_{SR}$ is simply the largest of $N_{T}M_{R}$ exponential RVs with parameter $\bar{\gamma}_{SR}$, $\gamma^{K,L}_{RD}$ is simply the largest of $M_{T}N_{R}$ exponential RVs with parameter $\bar{\gamma}_{RD}$, and, since the $R-R$ link is ignored, $\gamma^{I,L}_{R,R}$ is an exponential RV with parameter $\bar{\gamma}_{RR}$. The outage probability of MM AS can be written as
\begin{align}
\label{pmmeq3}
P_{\text{MM}}=1-\int^{\infty}_{0}\overline{F}_{X}\left(\frac{(y+\gamma_T+1)\gamma_{T}}{y}\right)f_{Y}(y+\gamma_T)dy,
\end{align}
where $X=\frac{\gamma^{I,J}_{SR}}{\gamma^{I,L}_{RR}+1}$, $Y=\gamma^{K,L}_{RD}$ and $\overline{F}_X\left(\cdot\right)$ denotes the complementary cdf of the RV, $X$. Clearly, in order to evaluate \eqref{pmmeq3} we first need to find the cdf and the pdf of $X$ and $Y$, respectively. The cdf of $X$ can be expressed as
\begin{align}
\label{mmeq2}
F_{X}(x) & = \frac{1}{\bar{\gamma}_{RR}}\int^{\infty}_{0}F_{\gamma^{I,J}_{SR}}\left((y+1)x\right)e^{-\frac{y}{\bar{\gamma}_{RR}}}dy\\\nonumber
& = 1-N_{T}M_{R}\sum^{N_{T}M_{R}-1}_{p=0}\frac{(-1)^p\binom{N_{T}M_{R}-1}{p}e^{-\frac{(p+1)x}{\bar{\gamma}_{SR}}}}{(p+1)\left(1+\frac{(p+1)\bar{\gamma}_{RR}x}{\bar{\gamma}_{SR}}\right)}.
\end{align}
The second equality in \eqref{mmeq2} follows since the binomial expansion $F_{\gamma^{I,J}_{SR}}\left(x\right)=\left(1-e^{-\frac{x}{\bar{\gamma}_{SR}}}\right)^{N_{T}M_{R}}$ can be written as $F_{\gamma^{I,J}_{SR}}\left(x\right)=1-N_{T}M_{R}\sum^{N_{T}M_{R}-1}_{p=0}\frac{(-1)^p \binom{N_{T}M_{R}-1}{p}}{p+1}e^{-\frac{(p+1)x}{\bar{\gamma}_{SR}}}.$
We can now write \eqref{pmmeq3} as
\begin{align}
\label{MMe1}
P_{\text{MM}}&=1-N_{T}M_{R}M_{T}N_{R}\sum^{N_{T}M_{R}-1}_{p=0}\frac{(-1)^p\binom{N_{T}M_{R}-1}{p}}{p+1}\sum^{M_{T}N_{R}-1}_{q=0}\frac{(-1)^q \binom{M_{T}N_{R}-1}{q}}{\bar{\gamma}_{RD}}\\\nonumber & \times  \int^{\infty}_{0}\frac{e^{-\frac{(p+1)(y+\gamma_T+1)\gamma_{T}}{\bar{\gamma}_{SR}y}-\frac{(q+1)(y+\gamma_T)}{\bar{\gamma}_{RD}}}}{\left(1+\frac{(p+1)(y+\gamma_T+1)\bar{\gamma}_{RR}\gamma_{T}}{\bar{\gamma}_{SR}y}\right)}dy.
\end{align}
Eq. \eqref{MMe1} does not admit a closed-form solution. However, it can be easily evaluated numerically using standard mathematical software tools.

In order to derive an accurate closed-form outage expression applicable in the asymptotic regime $(P_S \rightarrow \infty)$, we consider
\begin{align}
\label{HSNR1}
P_{\text{MM}} & \geq \mathsf{Pr}\left\{ \min \left( \frac{\gamma^{I,J}_{SR}}{\gamma^{I,K}_{RR}+1}, \gamma^{K,L}_{RD} \right) <\gamma_T\right\}\\\nonumber
& \rightarrow 1-N_{T}M_{R}M_{T}N_{R}\sum^{N_{T}M_{R}-1}_{p=0}\frac{(-1)^p\binom{N_{T}M_{R}-1}{p}}{(p+1)\left(1+\frac{(p+1)\bar{\gamma}_{RR}\gamma_T}{\bar{\gamma}_{SR}}\right)}\sum^{M_{T}N_{R}-1}_{q=0}\frac{(-1)^q \binom{M_{T}N_{R}-1}{q}}{q+1}e^{-\left(\frac{p+1}{\bar{\gamma}_{SR}}+\frac{q+1}{\bar{\gamma}_{RD}}\right)\gamma_T},
\end{align}
where $F_{\min\left(X,Y\right)}(\cdot)=1-\left(1-F_{X}\left(\cdot\right)\right)\left(1-F_{Y}\left(\cdot\right)\right)$ has been used. Ignoring the product term $F_{X}\left(\cdot\right)F_{Y}\left(\cdot\right)$ as it gives higher order terms, we observe that the asymptotic behavior of $P_{\text{MM}}$ can be further approximated as $P_{\text{MM}} \approx F_{X}\left(\gamma_T\right)+ F_{Y}\left(\gamma_T\right)$. Consider $F_{X}\left(\gamma_T\right)$ as $P_{S} \rightarrow \infty$; for small $x=\frac{\gamma_T}{\bar{\gamma}_{SR}}$ we can simplify $F_{X}\left(x\right) = \frac{e^{-\frac{x}{\bar{\gamma}_{SR}}}}{\bar{\gamma}_{RR}}\int^{\infty}_{0}\left(1-e^{-\frac{x}{\bar{\gamma}_{SR}}}\right)^{N_{T}M_{R}} e^{-\frac{y}{\bar{\gamma}_{RR}}}dy$ as
\begin{align}
F_{X}(x) & \approx \frac{x^{N_{T}M_{R}}}{\bar{\gamma}_{RR}} \int^{\infty}_{0} y^{N_{T}M_{R}}e^{-\frac{y}{\bar{\gamma}_{RR}}}dy \\\nonumber
& = \left(N_{T}M_{R}\right)!\left(\bar{\gamma}_{RR}x\right)^{N_{T}M_{R}}.
\end{align}
Similarly, we can show that as $P_S \rightarrow \infty$, $F_{Y}(\gamma_T)\approx  \left(\frac{\gamma_T}{\bar{\gamma}_{RD}}\right)^{M_{T}N_{R}}$. Therefore, \eqref{HSNR1} can be simplified for $0<\alpha < 1$ as
\begin{align}
\label{mmHsnr2}
P_{\text{MM}} & \approx \left(N_{T}M_{R}\right)!\left(\frac{\bar{\gamma}_{RR}\gamma_T}{\bar{\gamma}_{SR}}\right)^{N_{T}M_{R}}+\left(\frac{\gamma_T}{\bar{\gamma}_{RD}}\right)^{M_{T}N_{R}}.
\end{align}
As an immediate observation, from \eqref{OSHSNR} and \eqref{mmHsnr2} we see that the OP AS scheme and the MM AS scheme achieve the same diversity performance. As a result, we have $\alpha^{\text{MM}}_{\mathsf{opt}} = \alpha^{\text{OP}}_{\mathsf{opt}} $ with $P^{\alpha^{\text{MM}}_{\mathsf{opt}}}_{S}$ as the optimal power allocation solution at the relay and the highest diversity order, achieved with the MM AS scheme is also $d_{\text{max,MM}}=\frac{1}{\left(M_{T}N_{R}\right)^{-1}+\left(N_{T}M_{R}\right)^{-1}}$. However, compared to the MM AS scheme, the OP AS scheme has a higher array gain as verified in Section \ref{NUM_SEC}.

\subsubsection{Partial Antenna Selection}
The outage probability of this scheme can be evaluated from
\begin{align}
P_{\text{PR}} = 1-\int^{\infty}_{0}\overline{F}_{X}\left(\frac{(y+\gamma_T+1)\gamma_{T}}{y}\right)f_{Y}(y+\gamma_T)dy,
\end{align}
with $X=\frac{\gamma^{I,J}_{SR}}{\gamma^{I,L}_{RR}+1}$ and $Y=\gamma^{K,L}_{RD}$. The required distributions of $X$ and $Y$ are different to the previous case of $\max-\max$ AS and in order to calculate $P_{\text{PR}}$ we need to evaluate them. For any $i$-th relay receive antenna, the ratio $\frac{\gamma^{i,j}_{SR}}{\gamma^{i,l}_{RR}+1}$ is maximized when the strongest $S-R$ channel and the weakest $R-R$ channel from the $i$th antenna ($i=1,\ldots,M_{R}$) are selected. Since there are $M_{R}$ antennas, the cdf of $X$ can be evaluated as $F_{X}(x) = \left( \int^{\infty}_{0} F_{A}\left((y+1)x\right)f_B(y)dy \right)^{M_{R}}$, where $A$ is a RV defined as the largest among $N_{T}$ exponentially distributed RVs, while $B$ is the smallest out of $M_{T}$ exponentially distributed RVs. Substituting the required cdf and the pdf into $F_{X}(x)$ with simplifications yields
\begin{align}
F_{X}(x)= \left( 1 - N_{T}\sum^{N_{T}-1}_{p=0}\frac{(-1)^p\binom{N_{T}-1}{p}e^{-\frac{(p+1)x}{\bar{\gamma}_{SR}}}}{(p+1)\left(1+\frac{(p+1)\bar{\gamma}_{RR}x}{M_{T}\bar{\gamma}_{SR}}\right)} \right)^{M_{R}}.
\end{align}
Furthermore, we notice that the RV, $Y=\gamma^{K,L}_{RD}$, is simply the largest among $N_{R}$ exponential RVs with parameter $\bar{\gamma}_{RD}$. Therefore, the pdf of $Y$ can be written as $f_{Y}(y)=\frac{N_{R}}{\bar{\gamma}_{RD}}\sum^{N_{R}-1}_{q=0}(-1)^q \binom{N_{R}-1}{q}e^{-\frac{(q+1)y}{\bar{\gamma}_{RD}}}$. Combining these results, the exact outage probability of the PR AS scheme can be written as
\begin{align}
\label{intg}
P_{\text{PR}}=1-\frac{N_{R}}{\bar{\gamma}_{RD}}\sum^{N_{R}-1}_{q=0}(-1)^q\binom{N_{R}-1}{q}\mathcal{I}_q,
\end{align}
where the integral $\mathcal{I}_q$ is defined as
\begin{align}
\label{intexpr}
\mathcal{I}_q = \int^{\infty}_{0}\left(1-\left( 1 - N_{T}\sum^{N_{T}-1}_{p=0}\frac{(-1)^p\binom{N_{T}-1}{p}e^{-\frac{(p+1)(y+\gamma_T+1)\gamma_T}{\bar{\gamma}_{SR}y}}}{(p+1)\left(1+\frac{(p+1)\bar{\gamma}_{RR}(y+\gamma_T+1)\gamma_T}{M_{T}\bar{\gamma}_{SR}y}\right)} \right)^{M_{R}}\right)e^{-\frac{(q+1)(y+\gamma_T)}{\bar{\gamma}_{RD}}}dy.
\end{align}
In order to derive an accurate closed-form expression for the outage probability with $P_S \rightarrow \infty$ we consider
\begin{align}
\label{HSNRPS}
P_{\text{PR}} & \geq \mathsf{Pr}\left\{ \min \left( \frac{\gamma^{I,J}_{SR}}{\gamma^{I,K}_{RR}+1}, \gamma^{K,L}_{RD} \right) <\gamma_T\right\}\\\nonumber
& \approx \left(\frac{N_{T}!}{M_{T}^{N_{T}}}\right)^{M_{R}}\left(\frac{\bar{\gamma}_{RR} \gamma_T}{\bar{\gamma}_{SR}}\right)^{N_{T}M_{R}}+\left(\frac{\gamma_T}{\bar{\gamma}_{RD}}\right)^{N_{R}},
\end{align}
for $0<\alpha < 1$. We see that the first term decays as $P_S^{-(1-\alpha)N_{T}M_{R}}$ while the second term decays as $P_S^{-\alpha N_{R}}$. Therefore, outage minimization occurs when $(1-\alpha)N_{T}M_{R}=\alpha N_{R}$ and we have
\begin{align}
\alpha^{\text{PR}}_{\mathsf{opt}}=\frac{N_{T}M_{R}}{N_{T}M_{R}+N_{R}},
\end{align}
with $P^{\alpha^{\text{PR}}_{\mathsf{opt}}}_{S}$ as the optimal power allocation solution at the relay. Therefore, the highest diversity order, $d_{\text{max,PR}}$, achieved with the PR AS scheme can be expressed as
\begin{align}
d_{\text{max,PR}}=\frac{1}{N_{R}^{-1}+\left(N_{T}M_{R}\right)^{-1}}.
\end{align}

\subsubsection{Loop Interference Antenna Selection}
In the case of the LI AS scheme, the outage probability can be evaluated from
\begin{align}
P_{\text{LI}} = 1-\int^{\infty}_{0}\overline{F}_{X}\left(\frac{(y+\gamma_T+1)\gamma_{T}}{y}\right)f_{Y}(y+\gamma_T)dy,
\end{align}
where $X=\frac{\gamma^{I,J}_{SR}}{\gamma^{I,L}_{RR}+1}$ and $Y=\gamma^{K,L}_{RD}$. Since receive/transmit antennas at $R$ are selected to minimize the LI, with this scheme $\gamma^{I,L}_{RR}$ is the minimum of $M_{R}M_{T}$ exponential RVs with parameter $\bar{\gamma}_{RR}$, while $\gamma^{I,J}_{SR}$ and $\gamma^{K,L}_{RD}$ are the largest of $N_{T}$ and $N_{R}$ exponential RVs with parameters $\bar{\gamma}_{SR}$ and $\bar{\gamma}_{RD}$, respectively. Therefore, the required cdf of $X$ can be found using
\begin{align}
\label{LIeq2}
F_{X}(x) = 1 & - \frac{N_{T}M_{R}M_{T}}{\bar{\gamma}_{RR}} \sum^{N_{T}-1}_{p=0}\frac{(-1)^p \binom{N_{T}-1}{p}}{p+1} \int^{\infty}_{0} e^{-\frac{(p+1)(y+1)x}{\bar{\gamma}_{SR}}}e^{-\frac{M_{R}M_{T}y}{\bar{\gamma}_{RR}}}dy.
\end{align}
Simplifying the integral in \eqref{LIeq2} yields
\begin{align}
\label{LIAS1}
F_{X}(x) = 1 - N_{T}& \sum^{N_{T}-1}_{p=0}\frac{(-1)^p \binom{N_{T}-1}{p}e^{-\frac{(p+1)x}{\bar{\gamma}_{SR}}}}{(p+1)\left(1+\frac{(p+1)\bar{\gamma}_{RR}x}{M_{R}M_{T}\bar{\gamma}_{SR}}\right)}.
\end{align}
Now, combining the pdf of $Y$ and \eqref{LIAS1} we can express the exact outage probability as
\begin{align}
\label{LIexactres}
P_{\text{LI}} = 1 &-\frac{N_{T}N_{R}}{\bar{\gamma}_{RD}}\sum^{N_{T}-1}_{p=0}\frac{(-1)^p \binom{N_{T}-1}{p}}{p+1}\sum^{N_{R}-1}_{q=0}(-1)^q \binom{N_{R}-1}{q}\int^{\infty}_{0}\frac{e^{-\frac{(p+1)(y+\gamma_T+1)\gamma_T}{\bar{\gamma}_{SR}y}}e^{-\frac{(q+1)(y+\gamma_T)}{\bar{\gamma}_{RD}}}}{\left(1+\frac{(p+1)\bar{\gamma}_{RR}\gamma_T(y+\gamma_T+1)}{M_{R}M_{T}\bar{\gamma}_{SR}y}\right)}dy.
\end{align}
We now present an asymptotic approximation for the outage probability of the LI AS scheme. The outage probability as $P_S \rightarrow \infty$ can be approximated by
\begin{align}
\label{LIHSNR3}
P_{\text{LI}} & \geq \mathsf{Pr}\left\{ \min \left( \frac{\gamma^{I,J}_{SR}}{\gamma^{I,K}_{RR}+1}, \gamma^{K,L}_{RD} \right) <\gamma_T\right\}\\\nonumber
& \rightarrow 1 - N_{T}N_{R}\sum^{N_{T}-1}_{p=0}\frac{(-1)^p \binom{N_{T}-1}{p}e^{-\frac{(p+1)x}{\bar{\gamma}_{SR}}}}{(p+1)\left(1+\frac{(p+1)\bar{\gamma}_{RR}x}{M_{R}M_{T}\bar{\gamma}_{SR}}\right)}\sum^{N_{R}-1}_{q=0}\frac{(-1)^q \binom{N_{R}-1}{q}}{q+1}e^{-\frac{(q+1)y}{\bar{\gamma}_{RD}}}.
\end{align}
Eq. \eqref{LIHSNR3} can be simplified as
\begin{align}
\label{LIHS}
P_{\text{LI}} \approx \frac{N_{T}!}{(M_{R}M_{T})^{N_{T}}}\left(\frac{\bar{\gamma}_{RR}\gamma_T}{\bar{\gamma}_{SR}}\right)^{N_{T}}+\left(\frac{\gamma_T}{\bar{\gamma}_{RD}}\right)^{N_{R}},
\end{align}
for $0<\alpha < 1$. We see that the first term decays as $P_S^{-(1-\alpha)N_{T}}$ while the second term decays as $P_S^{-\alpha N_{R}}$. As for the previous AS schemes, the optimum $\alpha$ value can be found from $(1-\alpha)N_{T}=\alpha N_{R}$ and is given by
\begin{align}
\alpha^{\text{LI}}_{\mathsf{opt}}=\frac{N_{T}}{ N_{T} + N_{R}},
\end{align}
to yield $P^{\alpha^{\text{LI}}_{\mathsf{opt}}}_{S}$ as the optimal power allocation solution at the relay. 

Further, the highest diversity order, $d_{\text{max,LI}}$, achieved with the LI AS scheme can be expressed as
\begin{align}
d_{\text{max,LI}}=\frac{1}{N_{T}^{-1}+N_{R}^{-1}}.
\end{align}

\begin{table}[ptb]
\caption{Diversity order and complexity for the precoding designs and AS Schemes.}
\label{TABCSI}\vspace{0.1cm} \centering\renewcommand{\arraystretch}{1.6}
\begin{tabular}{|c|c|c|}
\hline
\textbf{Scheme} & \textbf{Diversity Order} & \textbf{Complexity} \\ \hline
\emph{Receive ZF} & $\min\left(N_{T}(M_{R}-1), M_{T}N_{R} \right)$ & high \\ \hline
\emph{Transmit ZF} & $\min\left(N_{T}M_{R}, \left(M_{T}-1\right)N_{R} \right)$ & high \\ \hline
\emph{OP AS} &  $\frac{1}{\left(M_{T}N_{R}\right)^{-1}+\left(N_{T}M_{R}\right)^{-1}}$ & $N_TM_RM_TN_R$  \\ \hline
\emph{MM AS} &  $\frac{1}{\left(M_{T}N_{R}\right)^{-1}+\left(N_{T}M_{R}\right)^{-1}}$ & $N_TM_R+M_TN_R$ \\ \hline
\emph{PR AS} & $\frac{1}{N_{R}^{-1}+\left(N_{T}M_{R}\right)^{-1}}$ & $N_TM_RM_T+N_R$  \\ \hline
\emph{LI AS} & $\frac{1}{N_{T}^{-1}+N_{R}^{-1}}$ & $N_T+M_RM_T+N_R$ \\ \hline
\end{tabular}
\end{table}

\subsection{Comparisons of the Schemes}
Table 1 summarizes the diversity order achieved from the investigated schemes as well as their associated complexity. The first main observation is that the precoding designs outperform the AS schemes in terms of diversity gain. The utilization of all antenna elements mitigates the LI effects and ensures a diversity order that is dominated by the weakest relaying branch. We note that due to the received/transmitted ZF operation, one antenna element is reserved for spatial cancellation at the relay's input/output, respectively. On the other hand, OP AS and MM AS schemes achieve similar diversity performance and significantly outperform the PR AS and LI AS schemes. Another interesting observation is that the diversity order of the PR AS scheme does not depend on $M_T$. Similarly, in the case of LI AS the diversity order is independent of the number of relay antennas. By comparing the results in Column 2 of Table 1, it is easy to see that with $M_R,M_T>1$
\begin{align}
d_{\text{precoding}}>d_{\text{OP}}=d_{\text{MM}}>d_{\text{PR}}>d_{\text{LI}}.
\end{align}
As for the complexity, the precoding schemes utilize all the antennas and require a radio frequency chain for each antenna element. In addition, the computation of the  beamforming vectors involves demanding mathematical operations such as matrix  multiplication, matrix inversion and eigen-decomposition giving a general complexity of $O(n^3)$. Therefore, although ZF precoding designs achieve higher diversity performance, they are characterized by a higher complexity in comparison to AS schemes. The proposed AS schemes also correspond to different complexities and are appropriate for networks with different computational capabilities. In order to provide a simple comparison of their complexity, we use as a metric the number of channels that should be examined in order to apply each AS scheme. It is worth noting that each channel in most of the cases is associated with a feedback channel (and a training process) in a centralized implementation. The OP AS examines all the possible combinations and therefore corresponds to a high complexity equal to $N_T M_R M_T N_R$ channels. The MM AS scheme decouples the AS selection into two independent groups and therefore has a complexity of $(N_T M_R+M_TN_R)$ channels. The PR AS scheme decouples the $R-D$ link in the selection process and gives a complexity of $(N_T M_R M_T+N_R)$ channels. Finally, the LI AS scheme is based on the LI channel and thus has a complexity of $(N_T+M_R M_T+N_R)$ channels.

\section{Numerical Results}\label{NUM_SEC}
In this section, we give numerical examples for the outage probability of the proposed precoding and AS schemes. The simulation set-up follows the system model of Section II with $R_0=2$ BPCU, and $c_{SR}=c_{RD}=1$. Although we have considered a symmetric setup, i.e., $c_{SR} = c_{RD}$, the main observations shown for AS schemes in Figs. 4-6 are also valid for asymmetric setups, where $c_{SR} \neq c_{RD}$.

\subsection{Joint Precoding/Decoding Designs}
Fig. 2 shows the results for the receive ZF based precoding design with different antenna configurations. The specific values of $N_T,M_R,M_T,N_R$ for each antenna configuration are shown inside the figure labels as $(N_T,M_R,M_T,N_R)$ respectively. These results reveal several interesting observations useful for system designers. The achievable diversity orders of the considered configurations, given by $\min\left(N_T(M_R-1),M_T N_R\right)$, are $1,2$ and $3$, respectively. Therefore, although only one receive antenna is used at $D$, the performance can be improved by selecting appropriate design parameters at $S$ and $R$. This attribute of the system is useful under different conditions; e.g., when fixed infrastructure based relays are employed, they can be equipped with many antennas while user terminals that act as relays have space constraints, and here the source can be equipped with many antennas. We also observe that although $(2,2,2,1)$ and $(2,3,2,1)$ enjoys a diversity order of two, the latter has a superior performance as a result of higher array gain. The same observation can be seen when $(3,2,3,1)$ and $(2,3,3,1)$ are compared. In the first case, additional performance gain is obtained via increasing $M_R$ (also $(2,3,2,1)$ has one more total number of antennas compared to $(2,2,2,1)$). However, in the second case, while $(3,2,3,1)$ and $(2,3,3,1)$ have the same number of total antennas, swapping $N_S$ with $M_R$ improves the outage probability. For comparison, we have included results for half-duplex hop-by-hop beamforming~\cite{GAYAN2} with two configurations, namely $(2,2,1,1)$ and $(2,3,3,1)$ and $\gamma_T = 2^{2R_0}-1$. These results can be compared for example with $(2,2,1,1)$ full-duplex operation and refer to the so called ``RF chain preserved'' condition and the ``number of antenna preserved'' (at the relay) condition.
\begin{figure}[t]
\centering
\includegraphics[width=0.6\linewidth]{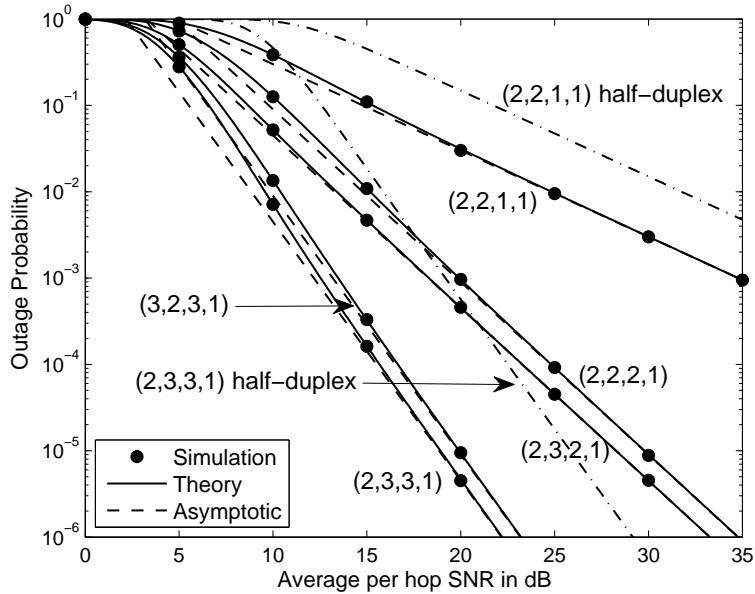}
\caption{Outage probability versus per hop average SNR for the receive ZF based precoding design with different antenna configurations.}
\label{FIG_RECEIVE_PRECODING}
\end{figure}

\begin{figure}[t]
\centering
\includegraphics[width=0.6\linewidth]{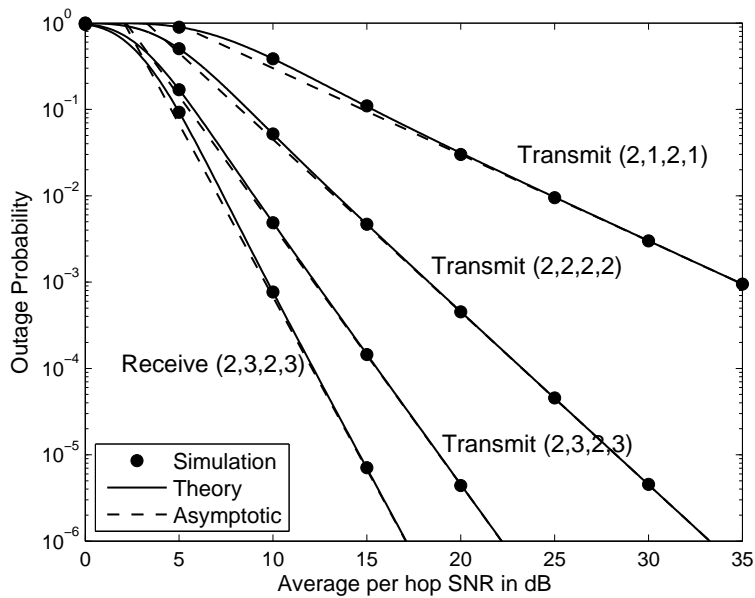}
\caption{Outage probability versus per hop average SNR of precoding designs with different antenna configurations.}
\label{FIG_TRANSMIT_PRECODING}
\vspace{-5mm}
\end{figure}

We show results for transmit ZF based precoding design with different antenna configurations in Fig. 3. The achievable diversity orders of the considered configurations, given by $\min\left(N_TM_R,(M_T-1) N_R\right)$, are again $1,2$ and $3$, respectively. We also compare the performance of the $(2,3,2,3)$ configuration under receive and transmit ZF designs, and the achievable diversity order of the former design given by $\min\left(N_T(M_R-1),M_T N_R\right)$ is four. Interestingly, receive ZF design exhibits a superior performance to transmit ZF since the former enjoys fourth order diversity order while the latter only has a diversity order of three. Clearly, this observation demonstrates that while under some configurations $(M_T=1)$ or $(M_R=1)$ only one form (receive or transmit) of precoding design can be deployed, in other configurations, when both designs can be applied, the system designer has to carefully decide on the configuration as well as the precoding design.

\subsection{Antenna Selection}

\begin{figure}[t]
\centering

\includegraphics[width=0.6\linewidth]{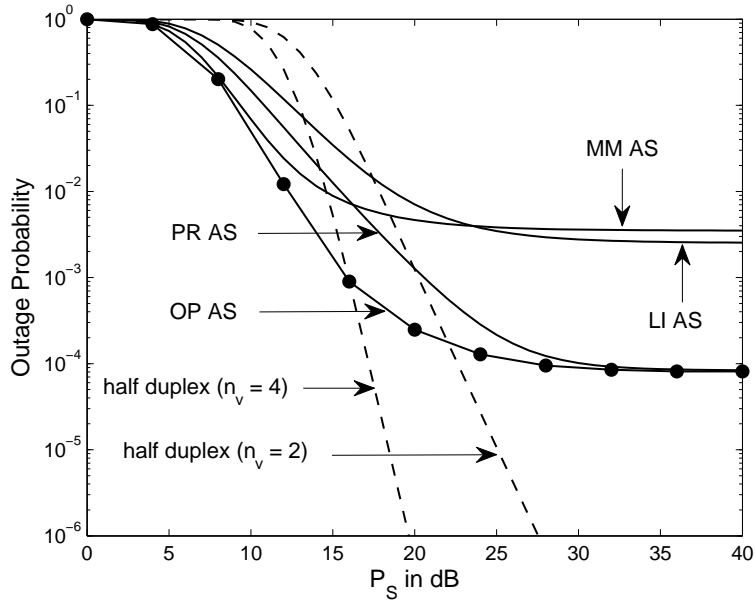}
\caption{Outage probability versus $P_S$; $c_{RR}=0.05$ and $\alpha=1$. The results for OP, MM, PR and LI AS schemes are computed via simulations and \eqref{MMe1}, \eqref{intg}, \eqref{LIexactres} respectively.}
\label{FIGG2}
\vspace{-5mm}
\end{figure}

\begin{figure}[t]
\centering
\includegraphics[width=0.6\linewidth]{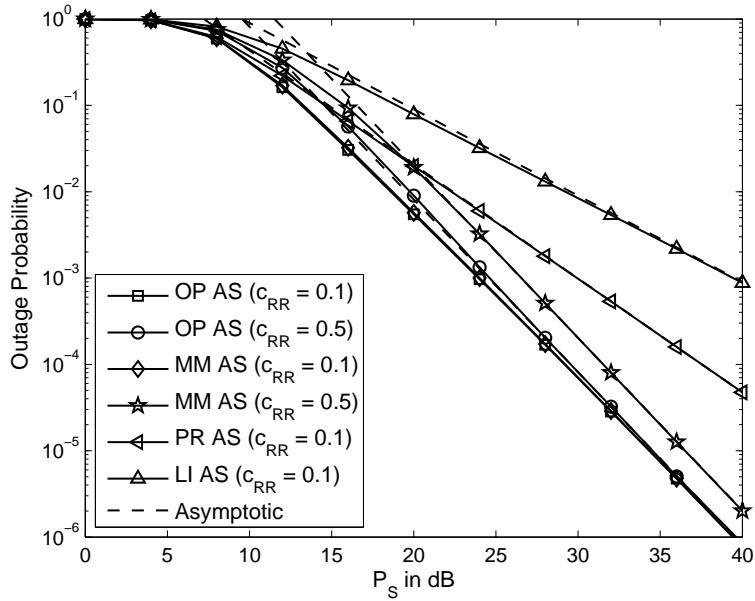}
\caption{Outage probability versus $P_S$ with optimum $\alpha$. The results for OP AS scheme are computed from simulations while asymptotic results for MM, PR, and LI AS schemes are due to \eqref{mmHsnr2}, \eqref{HSNRPS}, \eqref{LIHS} respectively.}\label{FIGG3}\vspace{-5mm}
\end{figure}

\begin{figure}[t]
\centering
\includegraphics[width=0.6\linewidth]{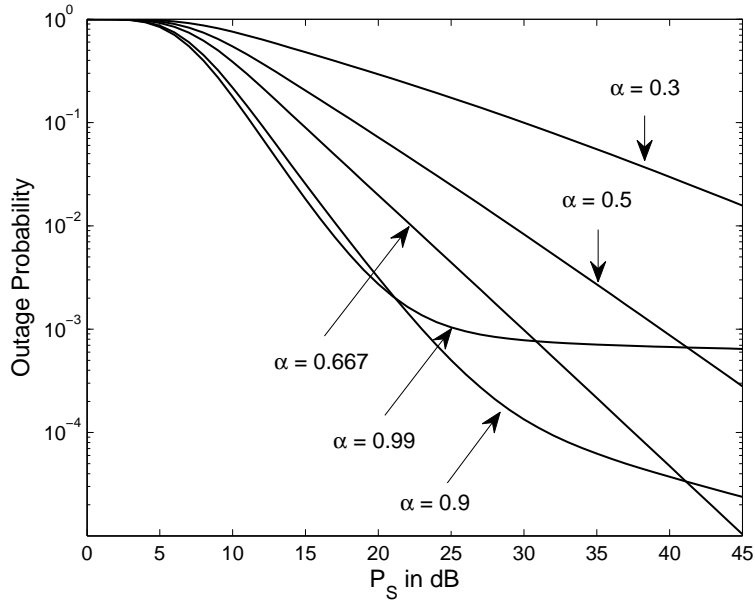}
\caption{Outage probability versus $P_S$ for the PR AS scheme and different $\alpha$; $c_{RR}=0.1$.}\label{fig:Figure4}\vspace{-5mm}
\end{figure}

In Figs.~\ref{FIGG2}-\ref{fig:Figure4}, we have set $N_{T}=M_{R}=M_{T}=N_{R}=2$. Fig. \ref{FIGG2} shows the outage probability as a function of $P_{S}$ for the considered AS schemes. No power control at $R$ is adopted and thus we adopt $\alpha=1$. Clearly, we see that all full-duplex schemes suffer from a zero-diversity order. Among the full-duplex AS schemes, the OP AS scheme provides the best performance. The PR AS scheme exhibits the next best performance and converges to the same error floor as the OP AS scheme. With low $P_S$, the MM AS performs better than both PR AS and LI AS schemes. Furthermore, for comparison with full-duplex, we have also plotted results for half-duplex operation with two cases; namely, the total number of antennas at the relay $(\textrm{n}_{V})$ is $2$ and $4$, respectively. With half-duplex transmission, the AS principle is simple; i.e., antennas are selected at each node to maximize the SNRs of the $S-R$ and $R-D$ links, respectively. The half-duplex results were plotted using \cite[Eq. (9)]{HWCNC10} with $\gamma_T = 2^{2R_0}-1$ due to the two time slot operation. The full-duplex AS schemes shows a favorable outage performance at a low-to-medium range of $P_S$, while the superiority of half-duplex transmission at high $P_S$ is clearly evident since it avoids LI and enjoys the benefits of diversity.

Fig. \ref{FIGG3} shows the outage probability of the AS schemes with optimal $\alpha$. In contrast to the results in Fig.~\ref{FIGG2}, where outage probability exhibits a saturated behavior at high $P_S$ (zero diversity), all AS schemes are now able to provide some diversity and outage decays as $P_S$ increases. For the considered system set up, $\alpha^{\text{OS}}_{\mathsf{opt}}=0.5, \alpha^{\text{MM}}_{\mathsf{opt}}=0.5, \alpha^{\text{PR}}_{\mathsf{opt}}=0.667$ and $ \alpha^{\text{LI}}_{\mathsf{opt}}=0.5$, and the achieved diversity orders of the OP, MM, PR and LI AS schemes are respectively, $2$, $2$, $1.33$ and $1$. Moreover, as expected, the OP AS scheme is able to provide the best performance among all the considered AS schemes in the work. When $c_{RR}$ is high $(0.5)$, a performance gap between OP AS and MM AS is observed (although both OP AS and MM AS provides the same diversity, the former has a higher array gain). However, we see that the performance difference between MM AS and OP AS schemes are almost negligible at $c_{RR}=0.1$. The usefulness of our asymptotic results can also be appreciated from Fig. \ref{FIGG3}. With increasing $P_S$, we see that the asymptotic plots match the exact results very well.

In Fig. \ref{fig:Figure4}, the outage behavior of the PR AS scheme with several values of $\alpha$ is illustrated. For $\alpha$ values close to one, the outage begins to suffer from low diversity (e.g., the curve corresponding to $\alpha=0.99$ almost converge to an error floor and exhibit a near zero diversity behavior). Clearly, the value of $\alpha^{\text{PR}}_{\mathsf{opt}}=0.667$ yields the best performance in the asymptotic regime. Interestingly, for $P_S<30$ dB, $\alpha=0.99$ and $0.9$ are able to provide a better performance than the optimal case before they begin to experience the decremental effects of low diversity. Therefore, depending on the operating region, an appropriate value for $\alpha$ can be selected. In the cases of OP AS, MM AS and LI AS, similar outage behavior with different $\alpha$ values can be observed as well.

\section{Conclusion}
In this paper, we considered full-duplex MIMO relaying with multi-antenna source and destination nodes. We introduced joint precoding/decoding designs which incorporate rank-1 zero-forcing self-interference suppression at the relay node. Our analysis delivered closed-form results which were further analyzed to reveal several interesting observations. Exact as well as asymptotic expressions for the outage probability were derived to explicitly reveal insights such as the achievable diversity order and the array gain. These results were also verified from simulations to confirm their correctness. The outage probability is influenced by the number of antennas deployed at each node as well as the adopted precoding (receive ZF or transmit ZF) design. In order to further reduce system complexity, we also presented several AS schemes. The investigated AS schemes have been analyzed in terms of the outage probability and exact expressions as well as asymptotic approximations have been derived. A simple power allocation scheme at the relay was proposed to overcome the zero-diversity limitation. A single parameter in the power allocation scheme can be set to obtain the desired outage performance while optimum values of this parameter were presented for diversity maximization of the investigated AS schemes.

\end{document}